\documentclass[prl,aps,twocolumn,showpacs,superscriptaddress,floatfix]{revtex4-1}

\usepackage{times}

\usepackage{graphicx}
\usepackage{amsmath}
\usepackage{amssymb}

\usepackage[utf8]{inputenc}
\usepackage[T1]{fontenc}

\newcommand{\Schro}{Schr\"odinger}
\newcommand{\SClabel}{\text{SC}}
\newcommand{\Llabel}{\text{L}}
\newcommand{\ad}{\text{ad}}
\newcommand{\nonad}{\text{nad}}
\newcommand{\texthom}{\text{hom}}
\newcommand{\DP}{\text{DP}}

\newcommand{\EPp}{\text{EP$+$}}

\newcommand{\EPpm}{\text{EP$\pm$}}
\newcommand{\CCW}{\text{CCW}}
\newcommand{\CW}{\text{CW}}
\newcommand{\ton}{\text{on}}
\newcommand{\toff}{\text{off}}

\newcommand{\idest}{\mbox{i.e.\ }}
\newcommand{\abs}[1]{\left| #1\right|}
\renewcommand{\Re}{\operatorname{Re}}
\renewcommand{\Im}{\operatorname{Im}}

\DeclareMathOperator{\Htheta}{\Theta}
\newcommand{\bigO}{\mathcal{O}}

\begin{document}

\title{Encircling exceptional points as a non-Hermitian extension of rapid adiabatic passage}

\author{J.\ \surname{Feilhauer}} \email{juraj.feilhauer@savba.sk}

\affiliation{Institute of Electrical Engineering, Slovak Academy of Sciences,
  841 04 Bratislava, Slovakia}
\affiliation{Institute for Theoretical Physics, Vienna University of Technology (TU-Wien), 1040 Vienna, Austria}

\author{A.\ \surname{Schumer}} 
\affiliation{Institute for Theoretical Physics, Vienna University of Technology (TU-Wien), 1040 Vienna, Austria}

\author{J. \surname{Doppler}} 
\affiliation{Institute for Theoretical Physics, Vienna University of Technology (TU-Wien), 1040 Vienna, Austria}

\author{A.\ A.\ \surname{Mailybaev}} 
\affiliation{Instituto Nacional de Matem\'atica Pura e Aplicada-IMPA, 22460-320 Rio de Janeiro, Brazil}
 
\author{J.\ \surname{B\"ohm}} 

\affiliation{Institut de Physique de Nice, Universit\'e C\^ote d'Azur, CNRS, 06108 Nice, France}

\author{U.\ \surname{Kuhl}} 

\affiliation{Institut de Physique de Nice, Universit\'e C\^ote d'Azur, CNRS, 06108 Nice, France}

\author{N.\ \surname{Moiseyev}} 

\affiliation{Schulich Faculty of Chemistry and Faculty of Physics, Technion-Israel Institute of Technology, 32000 Haifa, Israel}

\author{S.\ \surname{Rotter}}

\affiliation{Institute for Theoretical Physics, Vienna University of Technology (TU-Wien), 1040 Vienna, Austria}

\date{April 10, 2020}

\begin{abstract}
The efficient transfer of excitations between different levels of a quantum system is a task with many applications. Among the various protocols to carry out such a state transfer in driven systems, {\it rapid adiabatic passage} (RAP) is one of the most widely used. Here we show both theoretically and experimentally that adding a suitable amount of loss to the driven Hamiltonian turns a RAP protocol into a scheme for encircling an exceptional point including the chiral state transfer associated with it. Our work thus discloses an intimate connection between a whole body of literature on RAP and recent studies on the dynamics in the vicinity of an exceptional point, which we expect to serve as a bridge between the disjoint communities working on these two scenarios.
\end{abstract}

\maketitle
Already in the early years of quantum mechanics the question was discussed how to coherently transfer the population from one discrete energy level to another one \cite{landau1932,zener1932}. Ever since, coherent transfer schemes have become indispensable tools in many different areas of physics and technology -- from simple spin-flip operations in a magnetic field to the preparation of atoms with well-defined populations of their excited states \cite{allen2012optical}. A particularly efficient transfer scheme with an inherent robustness is ``rapid adiabatic passage'' (RAP) \cite{bloch1946nuclear,loy1974observation,vitanov2001laser,shore2008coherent,malinovsky2001general}.
While RAP is {\it rapid} enough to avoid any decoherence mechanism to kick in, it is  {\it adiabatic} in the sense that the external drive is smoothly turned on and off to keep the system in an instantaneous eigenstate of its time-dependent Hamiltonian. Since the Hamiltonian of the system must be the same before and after the external driving sequence, any initial state follows a closed loop in the space of driving parameters.
To achieve the desired population transfer, this closed loop must lead through an eigenvalue crossing at a degeneracy, also known as ``diabolic point’’ (DP) or ``conical intersection’’. 

In recent years, interest has been growing in non-Hermitian systems with controlled gain and loss that exhibit a wealth of unconventional and often surprising phenomena \cite{feng2017NatPhot,Ganainy2018NatPhys,Miri2019except,Ozdemir2019parity,longhi2018parity,bookNimrod}. The focus of the community’s attention currently revolves around the degeneracies that the corresponding non-Hermitian Hamiltonians give rise to. 
At these so-called ``exceptional points’’ (EPs) not only the complex eigenvalues of this effective Hamiltonian coincide (in both their real and imaginary parts), but also their corresponding eigenvectors become parallel \cite{bookKato,heiss2012physics,heiss2000repulsion,hahn2016observation,lee2009observation,gao2015observation,lefebvre2009resonance,atabek2011proposal,latinne1995laser,dembowski2001experimental}. 
Another key feature of non-Hermitian systems is their dynamics: even when the time evolution is arbitrarily slow, the adiabatic transport along the eigenvalue surfaces may break down at sudden non-adiabatic jumps during a state’s dynamical evolution \cite{uzdin2011observability,berry2011slow,berry2011optical,graefe2013breakdown,kapralova2014helium,PhysRevA.88.010102,milburn2015general,hassan2017chiral,hassan2017dynamically,choi2017extremely,ghosh2016exceptional}. One of the most surprising effects based on this breakdown of adiabaticity is a chiral state transfer: by continuously evolving two parameters along a closed loop around an EP, the final state at the end of the loop depends only on the encircling direction [clockwise (CW) vs.\ counterclockwise (CCW)], but not on the initial state. This interesting and robust effect has meanwhile been demonstrated in a number of different experimental platforms \cite{doppler2016dynamically,xu2016topological,Yoon2018time,zhang2018dynamically}. 

The  main  goal  of  this Letter is to demonstrate both theoretically and experimentally, that the above two transfer protocols are intimately connected in the sense that a RAP scheme in Hermitian systems results directly in the chiral encircling of an EP when losses are added appropriately.  
Our starting point is a general two-level system as described by the following simple $2\!\times\!2$ effective Hamiltonian, 
\begin{equation}
    \label{h2x2}
\mathcal{H} = \frac{1}{2}
    \begin{bmatrix}
          -\Delta - i \gamma & \Omega \\
              \Omega & \Delta + i \gamma
        \end{bmatrix},
\end{equation}
with the detuning $\Delta$, the coupling strength $\Omega$ and the loss or gain value $\gamma$ entering the two eigenvalues $\lambda_{\pm} = \pm \lambda$ and $\lambda = \sqrt{(\Delta + i \gamma)^2 + \Omega^2}/2\,$.
The dynamical evolution of the level amplitudes $c_1$ and $c_2$ is governed by the Schr\"odinger equation,
\begin{equation}
    \label{Schr}
i \frac{\partial}{\partial t}
\begin{pmatrix}c_1 \\ c_2 \end{pmatrix}
= \mathcal{H}(t) \begin{pmatrix}c_1\\c_2\end{pmatrix},
\end{equation}
where $\Delta$ and $\Omega$ in Eq.~(\ref{h2x2}) are smoothly adjusted with time.

\begin{figure}[tb!]
\centering
\includegraphics[clip,width=0.99\linewidth]{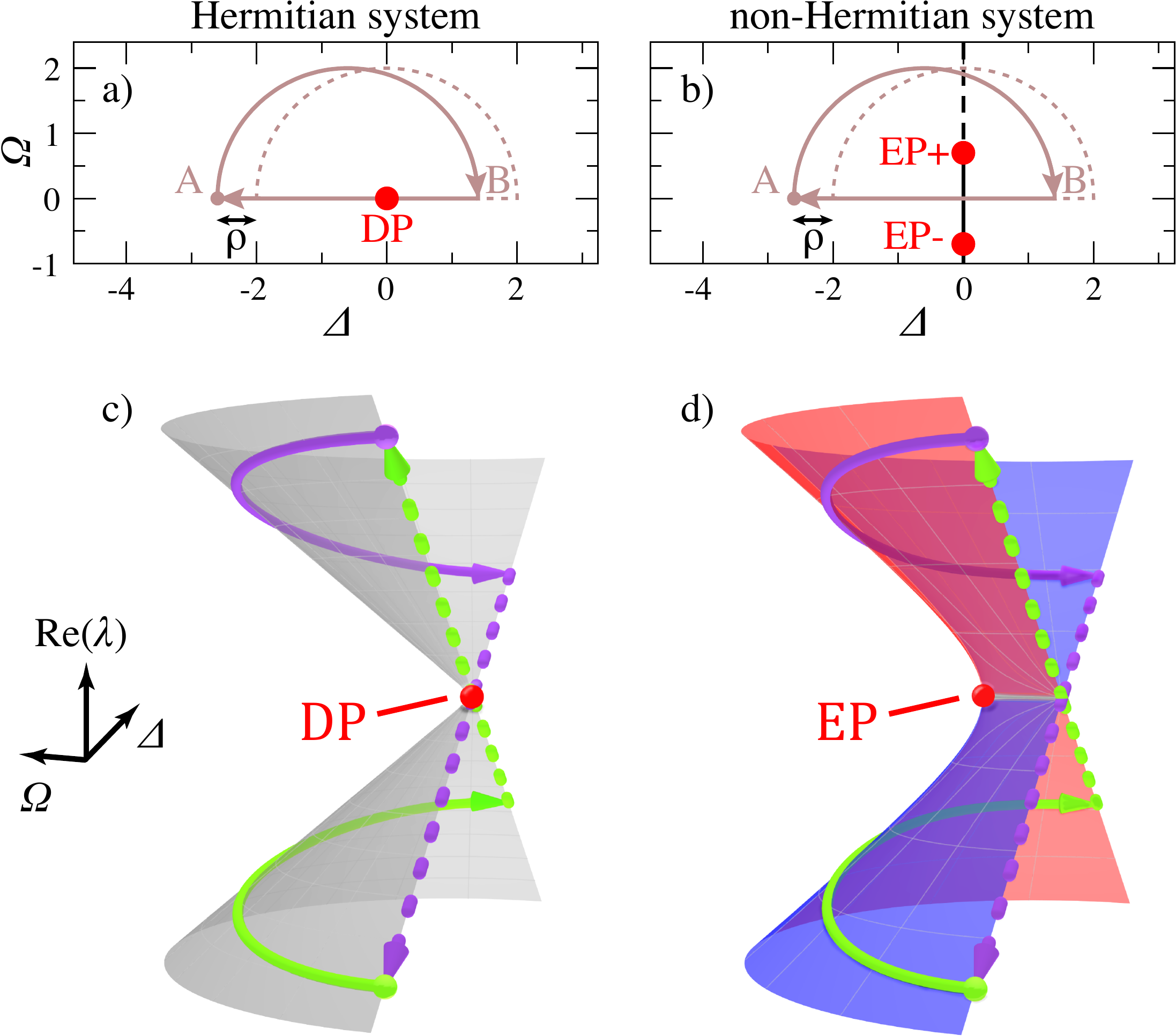}
\caption{(a, b) Clockwise adiabatic passage along a semicircular trajectory ($A\!\to\!B$) followed by a straight return path ($B\!\to\! A$) at zero coupling ($\Omega=0$). In the Hermitian case (a) the closed loop crosses a DP, whereas in the non-Hermitian case (b) it encircles one EP (labeled EP+). In both cases $\rho$ measures the loop offset from the center position at $\Delta=0$. In (c, d) we show the corresponding parametric state evolution by the arrows on the real part of the eigenvalue surfaces (for simplicity dynamical effects like non-adiabatic jumps are excluded here, see Fig.~\ref{Fig-2} for comparison).
Violet and green arrows show the state evolution starting on the first and second level, respectively. The red (gain) and blue (loss) regions in (d) correspond to $\text{Im} \lambda_{\pm} > 0$ and $\text{Im} \lambda_{\pm} < 0$, respectively.
} \label{Fig-1}
\end{figure}
In the Hermitian case without loss or gain ($\gamma = 0$) the \hyphenation{eigen-energy}eigenenergy surfaces $\lambda(\Delta,\Omega)$ 
in the parameter space ($\Delta,\Omega$) form a pair of conical sheets connected at a DP located at $\Delta=\Omega=0$ [see Fig.~\ref{Fig-1}(c)]. In Fig.~\ref{Fig-1}(a, c) we demonstrate
the adiabatic switch of the level populations associated with the RAP protocol by following a closed loop starting at point $A$ located at $\Omega = 0$. First we pass through a semicircular (SC) trajectory between points A and B ($SC_{A \rightarrow B}$) where the coupling $\Omega$ is switched on and off while the detuning $\Delta$ is simultaneously swept through the resonance at $\Delta = 0$. The closing of the loop
is achieved by connecting B with A along a linear (L) path ($L_{B \rightarrow A}$) by sweeping the detuning backwards at zero coupling through the DP. 
The important point to observe is that the desired final state at the end of the closed loop is realized already at point $B$ such that RAP usually terminates already there and the fictitious linear path  $L_{B \rightarrow A}$ is omitted (see SOM \cite{SOM} for more details). Since the same reasoning also holds when the semicircular loop is orbited in counterclockwise direction, the RAP protocol described above generates a symmetric eigenstate switch where any initial eigenstate is adiabatically exchanged at the end of the evolution regardless of the loop's orbiting direction. 
\begin{figure}[htb]
\centering
\includegraphics[clip,width=0.99\linewidth]{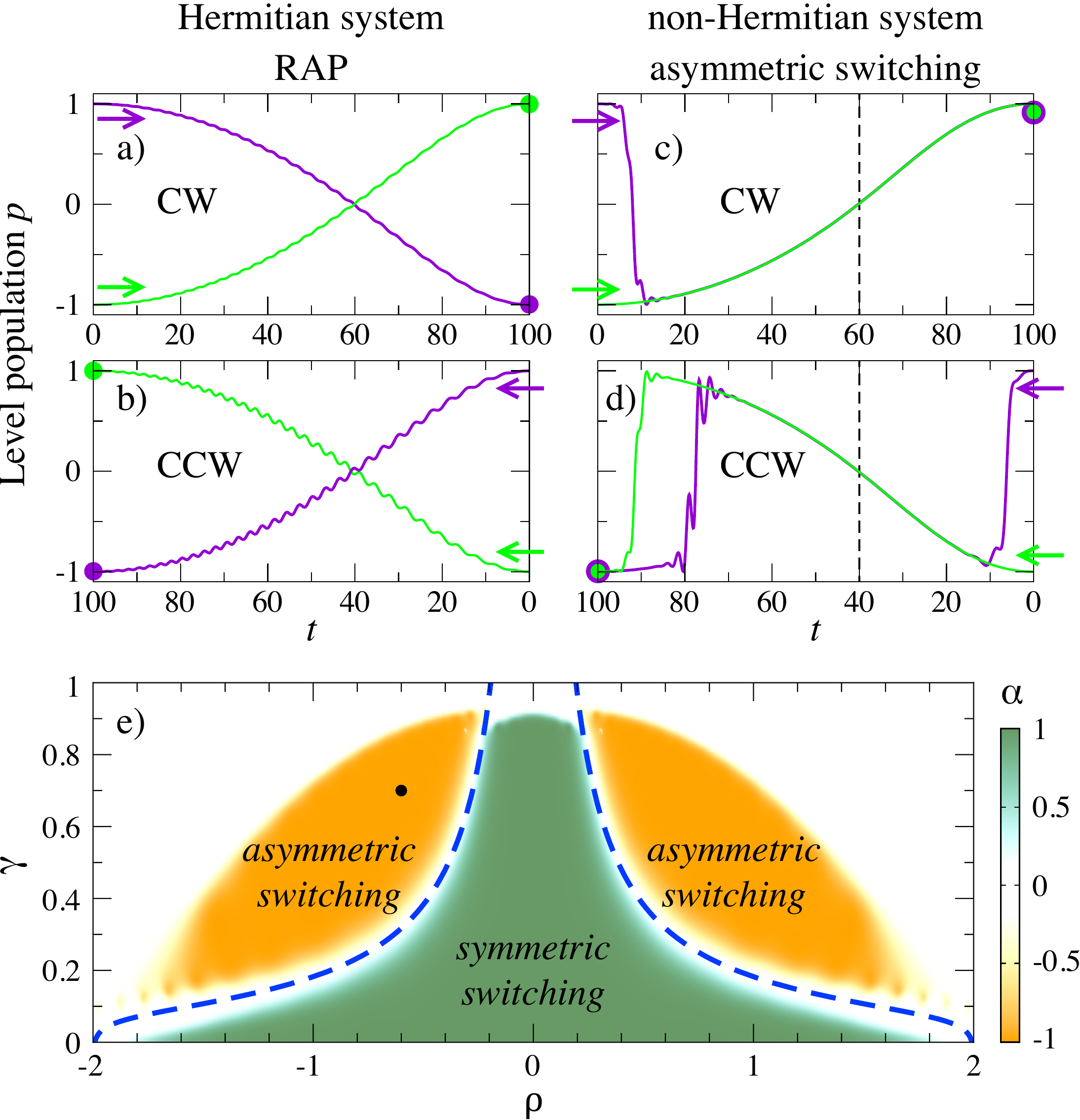}
\caption{(a)-(d) Dynamical evolution of the two level populations governed by Eq.~(\ref{Schr}) tracked along the semicircular path ($SC_{A \rightarrow B}$) in Fig.\ \ref{Fig-1} for the Hermitian (left panels) and non-Hermitian system (right panels, $\gamma = 0.7$). The numerical evolution of the level population $p$ is initialized (at $t=0$) in the first (violet curves) and second level (green curves). The arrows indicate the CW and CCW orbiting direction and the black dashed lines mark the position of the $\text{Im}\lambda = 0$ line. Violet and green dots mark the experimental values of mode populations $p_n$ from waveguide data in Fig.~\ref{Fig-4}. The bottom panel (e) shows a map of the state switch asymmetry when numerically following $SC_{A \rightarrow B}$ as a function of the loop offset $\rho$ and the loss-gain value $\gamma$. The shown switching parameter $\alpha$ takes on its limiting values $1$ ($-1$) for a symmetric (asymmetric) switch as in RAP (as in the chiral state transfer). Blue dashed line: transition from symmetric to asymmetric state transfer [see Eq.~(\ref{gamcSOM})]. Black dot: parameters of the semicircular loop in Fig.~\ref{Fig-1}(b).} \label{Fig-2}
\end{figure}

In a next step, we make the Hamiltonian in Eq.~(\ref{h2x2}) non-Hermitian by introducing a finite loss-gain value $\gamma > 0$ in its diagonal elements. With one level now being amplified and the other one being attenuated,  the eigenvalues $\lambda_\pm$ become complex with the real part representing the energy and the imaginary part defining the rate of amplification (for $\text{Im} \lambda_{\pm} > 0$) or attenuation (for $\text{Im} \lambda_{\pm} < 0$) of the corresponding eigenvectors [red and blue regions in Fig.~\ref{Fig-1}(d)]. The DP from the Hermitian case splits into a pair of two EPs located at $\Delta = 0$ and $\Omega_{\EPpm} = \pm \gamma$, where both the eigenvalues $\lambda_{\pm} = 0$ and the eigenvectors of Eq.~(\ref{h2x2}) coalesce. As shown in Fig.~\ref{Fig-1}(b), due to the symmetric splitting of the DP, one of the two EPs (at $\Omega_{\EPp} = \gamma$) is now located inside the closed parametric RAP loop whereas the second EP is not. The finite value of the loss-gain parameter $\gamma$ thus transforms the conventional RAP scheme into a protocol for EP-encircling. 

The topology of the energy eigensheets in the non-Hermitian system is shown in Fig.~\ref{Fig-1}(d). On these Riemann sheets a purely {\it parametric} evolution along the same semicircular RAP loop as in the Hermitian case also leads to the same symmetric state switch that RAP gives rise to [compare Fig.~\ref{Fig-1}(c) and (d) for the clockwise encircling direction]
\cite{bookNimrod,atabek2011proposal,dembowski2001experimental}. However, the fact that the {\it dynamic} state evolution of this non-Hermitian system is not necessarily adiabatic, leads to the asymmetric (chiral) state transfer associated with a loop around an EP \cite{berry2011slow,berry2011optical,graefe2013breakdown,uzdin2011observability,PhysRevA.88.010102,kapralova2014helium,milburn2015general,hassan2017chiral,hassan2017dynamically,doppler2016dynamically,xu2016topological,Yoon2018time,zhang2018dynamically,choi2017extremely,ghosh2016exceptional}. In this case, the final state of the evolution depends only on the encircling direction and is independent of the initial state. 

In Fig.~\ref{Fig-2}(a)-(d) we now provide the detailed numerical results for the fully {\it dynamical} state evolution for both the Hermitian and the non-Hermitian case. In both situations the level populations show the expected symmetric vs.\ asymmetric state exchange also when the fictitious straight line path across the resonance at $\Delta=\Omega=0$ is omitted and the evolution is restricted to just the semicircular RAP loop parametrized as
$\Omega_{\text{SC}}(t) = r \sin(\pi t/T)$ and $\Delta_{\text{SC}}(t) = \mp r \cos(\pi t/T) + \rho$.
We choose $r = 2$ as the radius of the semicircle, $\rho = -0.6$ as its horizontal offset, $T = 100$ as the  encircling period and the $\mp$ sign in $\Delta_{\text{SC}}(t)$ defines the CW and CCW orbiting direction. The inversion of the level population is described by the quantity
$p(t) = (|c_1(t)|^2 - |c_2(t)|^2)/(|c_1(t)|^2 + |c_2(t)|^2)$, where $p = \pm 1$ represent the cases where only the first or the second level is occupied, respectively. The violet curves in the top and middle row of Fig.~\ref{Fig-2} correspond to the evolution for the first level being initially populated [initial conditions $c_1(0) = 1$ and $c_2(0) = 0$] and the green curves for the second level initially populated [$c_1(0) = 0$ and $c_2(0) = 1$] . 

In Fig.~\ref{Fig-2}(a, b) (left column) one can clearly see the successful operation of RAP: the evolution along $SC_{A \rightarrow B}$ inverts the level populations for both orbiting directions and initial states (for more details on the conditions of adiabaticity see SOM \cite{SOM}). 
In Fig.~\ref{Fig-2}(c, d) (right column) the corresponding non-Hermitian evolution along the same path is depicted: 
one immediately sees that adiabaticity breaks down and non-adiabatic transitions arise that lead to a rapid transfer of populations towards the eigenvector with gain. 
The first such jump occurs for both orbiting directions at $t \approx 10$ [see violet curves in Fig.~\ref{Fig-2}(c, d)]. The ensuing evolution is almost identical for both initial states and orbiting  directions as it adiabatically follows the amplified instantaneous eigenvector (see also Fig.~S3 in the SOM \cite{SOM}). 
However, as the $\text{Im}\lambda = 0$ line (black dashed line) is crossed, the imaginary part of the eigenvalues changes sign. As a result, the state vector now follows the lossy eigenstate such that another non-adiabatic jump can occur. The onset of a potential non-adiabatic transition is, however, always delayed with respect to the sign change of $\text{Im}\lambda$  \cite{milburn2015general}, such that the remaining time of the loop decides whether or not the transition takes place. When the $\text{Im}\lambda = 0$ line is crossed asymmetrically in time for the two orbiting directions in a way that, e.g., the delay time is longer than the remaining loop time in CW direction [Fig.~\ref{Fig-2}(c)] but shorter than the remaining loop time in CCW direction [Fig.~\ref{Fig-2}(d)], the state transfer shows the characteristic chiral behavior. 
This observation explains why a finite offset $\rho$ of the semicircular loop with respect to the position of $\text{Im} \lambda = 0$ line (black dashed line) is required to induce this asymmetry.

To quantify the effect of the loop offset $\rho$ on the asymmetry of the state switch we introduce a switching parameter $\alpha \in [-1,1]$ (defined in the SOM \cite{SOM}) where a value of $1$ corresponds to symmetric switching (as in RAP) and $-1$ to the asymmetric switching (chiral state transfer). A map of $\alpha$ as a function of $\rho$ and $\gamma$ is plotted in Fig.~\ref{Fig-2}(e). The dark green region centered around $\rho = 0$ defines the parameters where the evolution along the semicircular loop yields a symmetric switch. 
Albeit symmetric, the evolution is here not necessarily adiabatic since two non-adiabatic jumps experienced during the EP encirclement would yield the same output eigenvector as the fully adiabatic passage [compare, e.g., violet curves in Fig.\ \ref{Fig-2}(b, d)]. The symmetric and fully adiabatic evolution (as in RAP) is possible only for $\gamma \ll 1$.       
As discussed above, introducing a sufficient asymmetry $\rho$ of the loop position with respect to the $\text{Im} \lambda = 0$ line leads to the asymmetric switching (orange regions).
Using the theory of stability loss delay \cite{milburn2015general} we estimate analytically the critical loss rate $\gamma_c$ needed to transition from symmetric to asymmetric switching for $\gamma \ll r$,  
\begin{equation}
    \label{gamc}
    \gamma_c \simeq \frac{2 r}{T \abs{\rho}} \ln \left[ \frac{2 T ( r^2 - \rho^2)}{\pi r} \right].
\end{equation}
This estimate, shown as a blue dashed line in Fig.~\ref{Fig-2}(e), accurately reproduces the numerically calculated crossover. 

Reconsidering experimental RAP protocols that are always slightly non-Hermitian due to inevitable losses to the environment, we may thus conclude that RAP is typically realized as an EP encirclement with a sufficiently small loss contrast between the eigenmodes ($\gamma \ll \gamma_c$). In turn, simply increasing the loss contrast ($\gamma > \gamma_c$) at a sufficient offset value $\abs{\rho}$ produces the chiral state transfer. For even larger values of $\gamma$ and $\abs{\rho}$ the evolving states eventually all yield $\alpha = 0$ [white region in Fig.~\ref{Fig-2}(e)] since they all collapse into the gain eigenstate, regardless of the initial configuration and orbiting direction.

\begin{figure}[htb]
\centering 
\includegraphics[clip,width=0.99\linewidth]{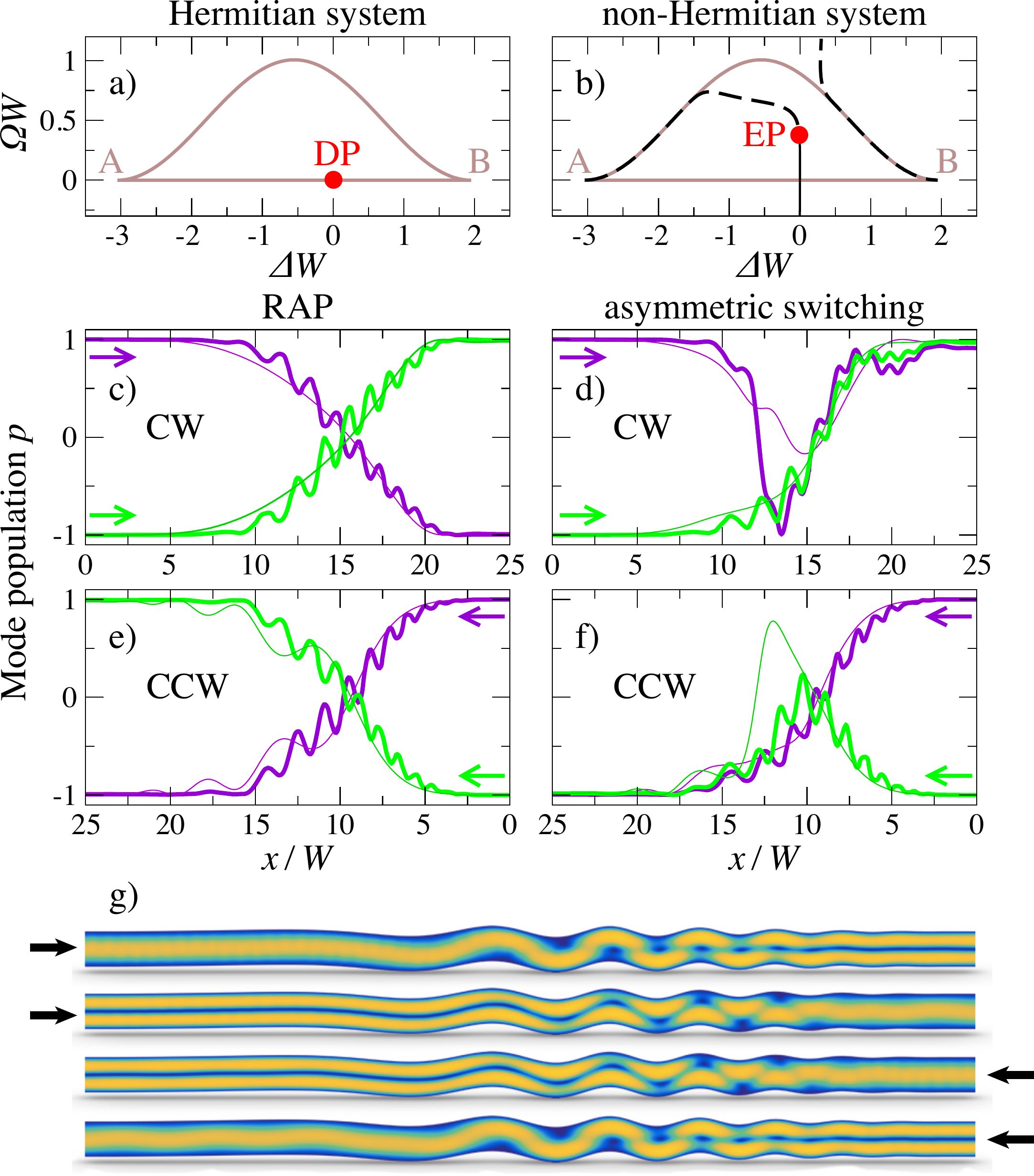}
\caption{Numerically calculated mode populations in the bimodal waveguide for $\omega W /\pi c = 2.6$ without (left column) and with absorber (right column). The top row shows the closed loops in parameter space which cross the DP in the case of an empty waveguide and enclose the EP in the case of a waveguide with absorber. Black dashed line in (b) marks the position of $\mbox{Im} \lambda = 0$ line. Panels (c)-(f) show the mode populations of microwaves injected into the waveguide from left and right (see arrows), calculated numerically (thick lines) and semi-analytically (thin lines). Panel (g) shows how the numerical field intensities of the transverse modes switch during the RAP protocol in the empty waveguide (arrows: injection).} \label{Fig-3}
\end{figure}

Our next aim is to test these theoretical considerations experimentally by implementing tunable losses in a system experiencing RAP in order to induce a chiral state transfer. Specifically, we choose for this purpose a bimodal  waveguide (WG) for microwaves in which the encircling of an EP and the associated chiral state transfer have recently been demonstrated \cite{doppler2016dynamically}. 
The transmission of microwaves through this bimodal waveguide of length $L$ and width $W$ can be modeled by the Schr\"odinger equation (\ref{Schr}) with the longitudinal coordinate $x$ playing the role of time. The coupling between the modes is provided by the simultaneously oscillating waveguide boundaries [see Fig.~\ref{Fig-3}(g) for an illustration] defined by the function $\xi(x) = \sigma \sin(k_b x)$ with amplitude $\sigma$ and $k_b = k_1 - k_2 + \delta$, where $\delta$ represents the detuning from the resonant forward scattering at $\delta = 0$. As shown in detail in \cite{doppler2016dynamically,SOM}, the modal amplitudes are governed by an $x$-dependent Hamiltonian similar to Eq.~(\ref{h2x2}), where the parameter  $\sigma$ is equivalent to $\Omega$ and $\delta$ is directly related to $\Delta$.
In order to reduce the backscattering of the microwaves at the start and end points $A,B$ we modify the semicircular RAP loop from Fig.~\ref{Fig-2} to the bell curve shown in Fig.~\ref{Fig-3}(a, b) with the corresponding waveguide boundary modulation given by $\sigma(x) = \sigma_0 [1 - \cos(2 \pi x/L)]/2$ and $\delta(x) = \pm \delta_0 (2 x/L - 1) + \rho$ for $\delta_0 W = 1.25$, $\rho W = -1.8$, $\sigma_0/W = 0.16$ and $L/W = 25$ \cite{doppler2016dynamically}.
The CW and CCW propagation along this loop is equivalent to the left and right injection of microwaves into the waveguide. 

The results of a full wave simulation \cite{rotter2000modular,libisch2012coherent} for the transmission across this bimodal waveguide are presented in Fig.~\ref{Fig-3} (see SOM \cite{SOM} for details). In the left column we consider the Hermitian case of this empty waveguide and observe a robust symmetric state switch for both initial modes and wave injection directions [see population inversions in Fig.~\ref{Fig-3}(c, e) (thick lines)]. This behavior is also well reproduced by a semi-analytical model based on Eq.~(\ref{Schr}) [see SOM and Fig.~\ref{Fig-3}(c, e) (thin lines)]. The successful implementation of RAP is visible also directly through the microwave intensity profiles along the boundary-modulated waveguide shown in Fig.~\ref{Fig-3}(g). 

In the non-Hermitian case, we consider a thin but strong absorber with suitable shape inside the waveguide to produce a sufficiently large loss-contrast between the two modes (see SOM \cite{SOM}). 
To identify the position of the EP with respect to the chosen parameter loop, we extend the definition of the waveguide 
Hamiltonian to the whole parameter plane, involving both the localized absorber and the homogeneous dissipation in the waveguide cover plates (see SOM \cite{SOM}). As shown in Fig.~\ref{Fig-3}(b), for the given waveguide geometry the EP is located inside the parametric loop and the evolution along the loop leads to the chiral mode switch for which the final states ejected at the two waveguide ends depend solely on the injection direction of the microwaves, but not on the injection profile, see Fig.~\ref{Fig-3}(d, f) \cite{doppler2016dynamically}. 

\begin{figure}[t]
\centering
\includegraphics[clip,width=0.99\linewidth]{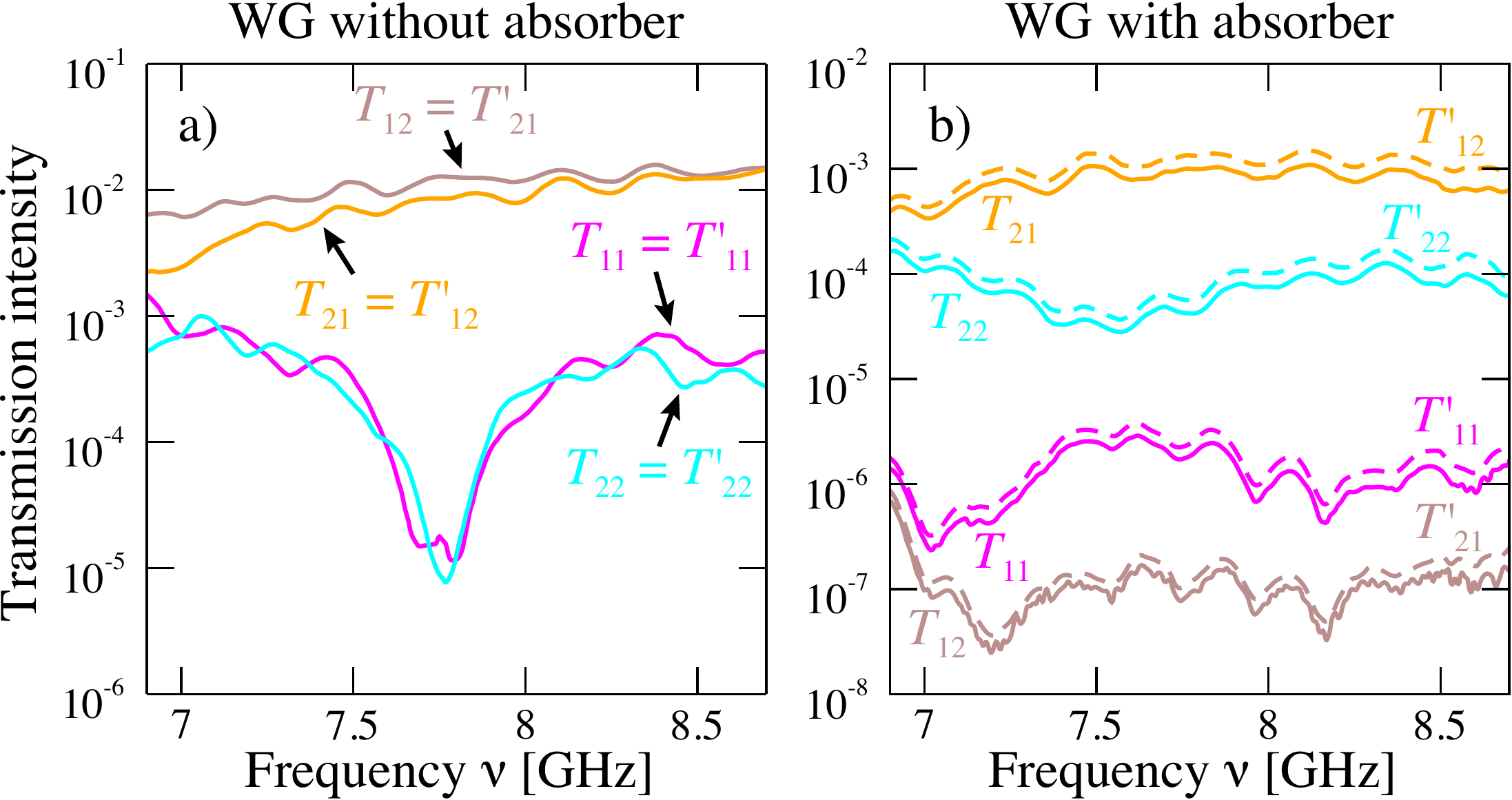}
\caption{Experimental transmission intensities in the microwave waveguide without (left panel) and with absorber (right panel). The solid and dashed curves correspond to the quantities of the left-to-right ($T_{mn}$) and right-to-left ($T'_{mn}$) transmission intensities from mode $m$ into mode $n$. Data in (b) reproduced from \cite{doppler2016dynamically}. 
} \label{Fig-4}
\end{figure}

Our theoretical results from above demonstrate that RAP and chiral state transfer can be obtained within the same waveguide. For switching between these two scenarios it is sufficient to adequately place an absorber into a waveguide that features RAP in the lossless case. We built such a \hyphenation{wave-guide}waveguide out of aluminum with dimensions $L \times W \times H = 2.38\ \text{m} \times 5\ \text{cm} \times 8\ \text{mm}$ consisting of a $1.25$ m long region with undulating boundaries between two straight waveguide leads \cite{doppler2016dynamically}. Microwaves with a frequency $\nu$ around $7.8$ GHz are injected and detected by antennas located in the leads. Fig.~\ref{Fig-4} shows the inter/intra-mode transmittances $T_{mn}$ (from left to right) and $T'_{mn}$ (from right to left) measured as a function of microwave frequency in the absence (left panel) and in the presence (right panel) of a thin foam absorber. In the waveguide without absorber, we observe that around the design frequency of $\nu = 7.8$ GHz the intermode transmittances are an order of magnitude larger than the intramode ones, which proves the successful operation of RAP.  
In the waveguide including the absorber the measured intensities satisfy $T_{11} \gg T_{12}$, $T_{21} \gg T_{22}$ and $T'_{12} \gg T'_{11}$, $T'_{22} \gg T'_{21}$, which is a hallmark of the chiral state transfer: both modes injected from the left (solid curves) leave the waveguide primarily in the first mode and both modes injected from the right (dashed curves) leave the waveguide primarily in the second mode. To compare these experimental results directly with our simple model from Eq.~(\ref{h2x2}), we mapped these transmittances at $\nu = 7.75$ GHz to corresponding mode populations via $p_n = (T_{n1} - T_{n2})/(T_{n1} + T_{n2})$, $n = 1,2$ and included them in Fig.~\ref{Fig-2} as violet (for $n = 1$) and green points (for $n = 2$) at the end of the corresponding loops. Our experimental values for $p_n$ nicely correspond to the final states of the evolution in the general model, confirming the successful experimental implementation of our theoretical concepts.

In summary, we have demonstrated the intimate connection between RAP in Hermitian systems and EP-encircling in non-Hermitian systems. Using analytical and numerical tools, we have shown explicitly that judiciously adding dissipative loss to a RAP protocol generates a chiral transfer scheme involving the encircling of an EP. These results were implemented in a boundary-modulated waveguide with a mode-specific absorber inside. In the absence of the absorber we observed the symmetric state switch of RAP between all incoming and outgoing modes. In the presence of the absorber we observed chiral transmission that depends primarily on the injection port (left or right), but not on the incoming mode configuration. \\
\begin{acknowledgments}
J.F.\ wishes to acknowledge the support of grant VEGA 2/0162/18, of the Action Austria-Slovakia and the National Scholarship Programme of the Slovak Republic. A.S.\ and S.R.\ were partly supported by the European Commission under project NHQWAVE No.\ MSCA-RISE 691209 and by the Austrian Science Fund (FWF) under project number P32300 (WAVELAND). N.M.\ acknowledges the support of the Israel Science Foundation Grant No.\ 1661/19. 
\end{acknowledgments}

\bibliographystyle{apsrev4-1}
\bibliography{EP_RAP_main_SOM.bib}

\onecolumngrid
\clearpage
\setcounter{equation}{0}
\setcounter{figure}{0}
\setcounter{table}{0}
\setcounter{page}{5}
\makeatletter
\renewcommand{\thefigure}{S\arabic{figure}}
\renewcommand\theequation{S.\arabic{equation}}
\renewcommand{\bibnumfmt}[1]{[S#1]}
\renewcommand{\citenumfont}[1]{S#1}

\begin{center}
    \textbf{\Large  Supporting Online Material:\\
Encircling exceptional points as a non-Hermitian extension \\
of rapid adiabatic passage}
\end{center}

\section*{Closed semicircular loop in the general two-level model}

In order to relate rapid adiabatic passage (RAP) and the chiral state transfer resulting from encircling of an exceptional point (EP) along a closed loop, we prove here that guiding the Hamiltonian of a system along an open semicircular trajectory such that it exhibits RAP is effectively equal to the corresponding closed semicircular loop. Of course, this only holds true if the semicircle is closed along the line defined as $\Omega = 0$ (no coupling).

In the following we only discuss the clockwise (CW) passage of the semicircular loop from the main text which consists of the semicircle $SC_{A \rightarrow B}$ and the straight line $L_{B \rightarrow A}$. A suitable parametrization for $SC_{A \rightarrow B}$ reads 
\begin{gather}
    \label{hcge}
  \Omega_{\SClabel}(t) = r \sin(\pi t/T_{\SClabel}), \\
    \label{hcom}
  \Delta_{\SClabel}(t) = -r \cos(\pi t/T_{\SClabel}) + \rho,
\end{gather}
with $0 \le t \le T_{\SClabel}$. The starting point $A$ and end point $B$ are located at $\Omega = 0$. The linear part $L_{B \rightarrow A}$ of the loop is defined as
\begin{gather}
    \label{linege}
  \Omega_{\Llabel}(t) = 0, \\
    \label{lineom}
  \Delta_{\Llabel}(t) = r [ 1 - 2 (t - T_{\SClabel})/T_{\Llabel} ] + \rho,
\end{gather}
with $T_{\SClabel} < t < T$, where $T$ is the total evolving time along the loop, $T = T_{\SClabel} + T_L$. 

As the time evolution along the straight line $L_{B \rightarrow A}$ occurs while the levels are decoupled (\idest $\Omega_L = 0$), we can write down the analytical solution of the \Schro\ equation 
\begin{align}
    \label{reslinea}
  c_1 (T) &= e^{i \int_{T_{\SClabel}}^{T} \Delta_L (t') dt'/2}\,e^{-\gamma T_{\Llabel}/2}\,c_1 (T_{\SClabel}), \\
    \label{reslineb}
  c_2 (T) &= e^{-i \int_{T_{\SClabel}}^{T} \Delta_L (t') dt'/2}\, e^{+\gamma T_{\Llabel}/2}\, c_2 (T_{\SClabel}),
\end{align}
where $c_i (T_{\SClabel})$ and $c_i (T)$, $i = 1,2$, are the level amplitudes after traversing the semicircle $SC_{A \rightarrow B}$ and the semicircular loop ($SC_{A \rightarrow B}$, $L_{B \rightarrow A}$), respectively.
In the Hermitian case, \idest for $\gamma = 0$, the linear part $L_{B \rightarrow A}$ necessary to close the loop only generates a phase factor. To eliminate the influence of the linear part in the general case ($\gamma \neq 0$), we let $T_L \rightarrow 0$. As the levels are perfectly decoupled along $L_{B \rightarrow A}$ this does not influence the overall adiabaticity of the loop. Then we get $T = T_{\SClabel}$ which is already used in Eqs.~(4) and (5) in the definitions of the semicircular loop in the main text.
The closing of the loop is therefore fictitious and it only serves the purpose of illustrating the accumulation of the geometric phase of the eigenvectors and the resulting swap with respect to the initial eigenbasis.
Equations~(\ref{hcge}--\ref{lineom}) define a semicircular loop traversed in CW direction. For the corresponding counterclockwise (CCW) encirclement, the first segment of the loop is the linear part $L_{A \rightarrow B}$ followed by the semicircle $SC_{B \rightarrow A}$. Based on the same reasoning we can neglect the linear part also for the CCW case. 

\section*{Adiabaticity in the Hermitian case}
\begin{figure}[t]
\centering 
\includegraphics[clip,width=0.55\linewidth]{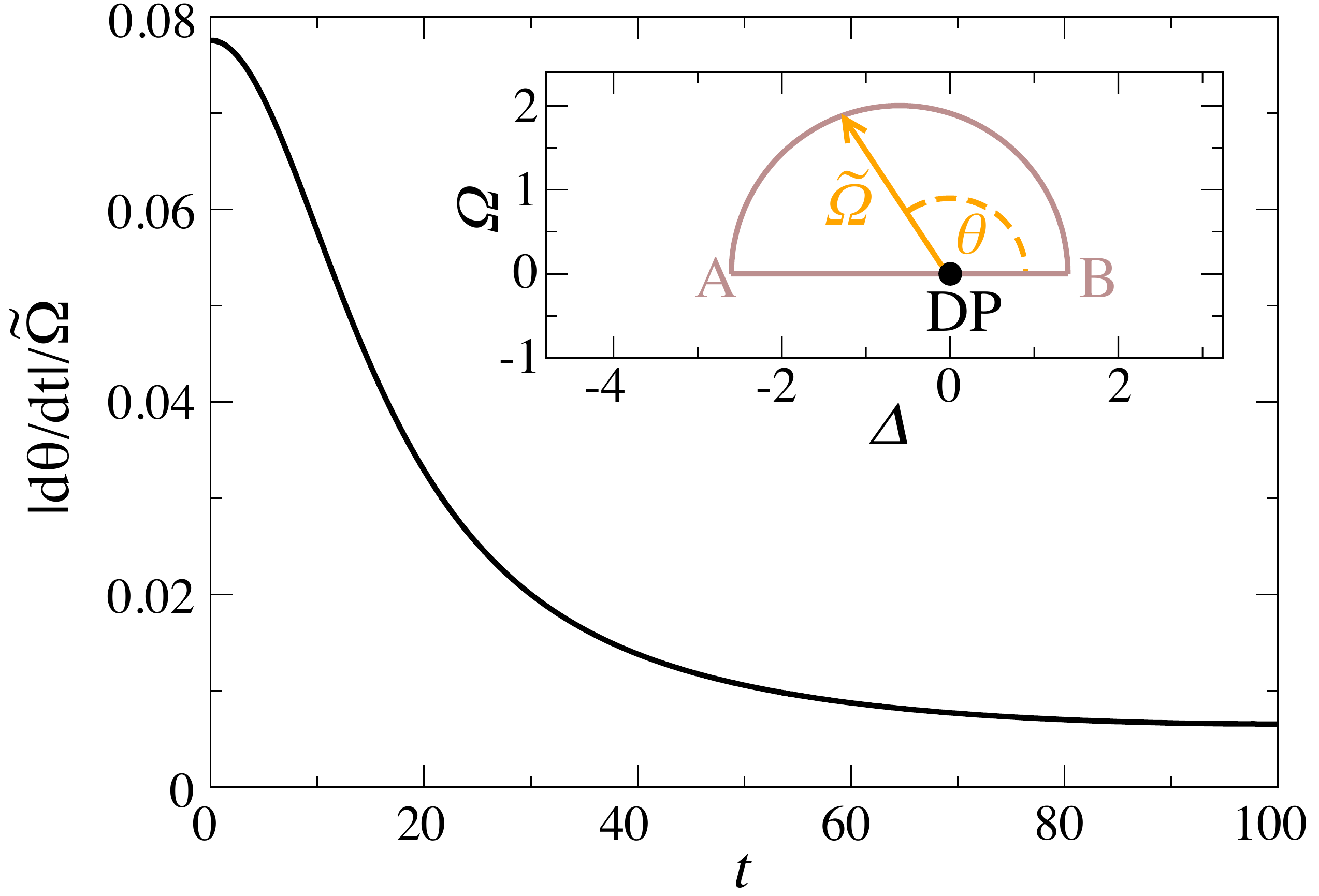}
\vspace{-0.0cm}
\caption{Tracking the adiabaticity condition $\abs{d \theta /dt}/ \tilde{\Omega} \ll 1$ when traversing the semicircular parameter loop displayed in Fig.~1(a) in the main text (see also inset). Clearly, the condition is perfectly satisfied in the simplified Hermitian model.} \label{S-1}
\end{figure}

As RAP relies on the state vector to adiabatically follow an eigenstate when the time evolution is governed by a Hermitian Hamiltonian
\begin{equation}
    \label{wgh2x2herm}
H_0(t) =
    \frac{1}{2}\begin{bmatrix}
          -\Delta(t) & \Omega(t) \\
               \Omega(t) & \Delta(t)
        \end{bmatrix},
\end{equation}
the parameter cycle should be carried out sufficiently slowly to avoid unwanted non-adiabatic population transfer. Adiabaticity is secured when during the time evolution driven by $H_0(t)$ the state vector $\vec{\psi}(t)$, initially prepared in an eigenstate $\vec{r}_{\pm}(0)$, remains close to the same instantaneous eigenstate $\vec{r}_\pm(t)$. This condition can be quantified \cite{malinovsky2001generalSOM} as
\begin{equation}
    \label{adia}
    \tilde{\Omega}(t) = \sqrt{ \Omega(t)^2 + \Delta(t)^2} \gg |d \theta(t)/dt|,
\end{equation}
where $\tilde{\Omega}$ represents the energy gap between the upper and lower eigenenergy sheets of Eq.~(\ref{wgh2x2herm}) for a given $\Omega$ and $\Delta$ while $\theta$ represents the phase angle. Moreover, $\tilde{\Omega}$ also equals the length of the vector $(\Delta,\Omega)$ [see inset of Fig.~\ref{S-1}] oriented at an angle $\theta$ that satisfies $\tan\theta = \Omega/\Delta$. Therefore, the graphical interpretation of the adiabaticity condition in Eq.~(\ref{adia}) states that during the adiabatic passage the length of the vector $\vec{\tilde{\Omega}}$ should be much larger than its angular velocity. For the semicircular part of the loop, the passage time $T$ and the radius $r$ can be tuned to satisfy the condition while for the linear part $\theta$ is constant and the evolution therefore perfectly adiabatic. The values of Eq.~(\ref{adia}) for the loop in  Fig.~2(a) in the main text are shown in Fig.~\ref{S-1}.
Obviously, the adiabaticity condition is perfectly satisfied throughout the entire passage along the semicircular loop.

\begin{figure}[t]
\centering 
\includegraphics[clip,width=0.55\linewidth]{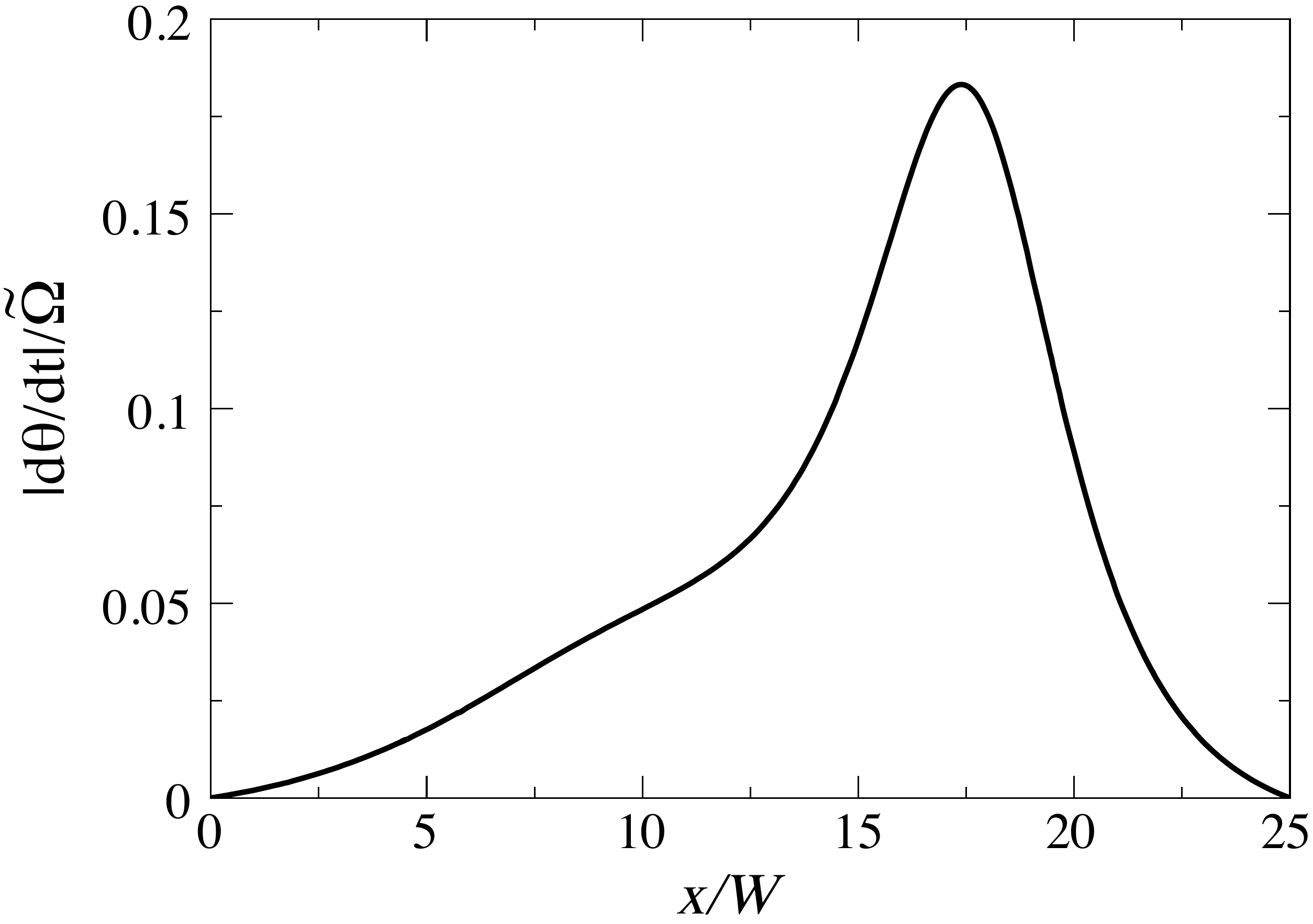}
\vspace{-0.0cm}
\caption{Tracking the adiabaticity condition $\abs{d \theta /dt}/\tilde{\Omega} \ll 1$ when passing the lossless waveguide according to the parametric loop shown in Fig.~3(a) of the main text. The condition is again satisfied. } \label{S-2}
\end{figure}

We can also test the adiabaticity condition for wave transport in the finite bimodal waveguide without absorber when considering the Hermitian part of the Hamiltonian [see Eq.~(\ref{wgh2x2})]. The obtained values of $\abs{d \theta /dt}/\tilde{\Omega}$ shown in Fig.~\ref{S-2} confirm that the adiabaticity condition is satisfied also for the simulation of the microwave waveguide.

\begin{figure}[!ht]
\centering
\includegraphics[clip,width=0.6\linewidth]{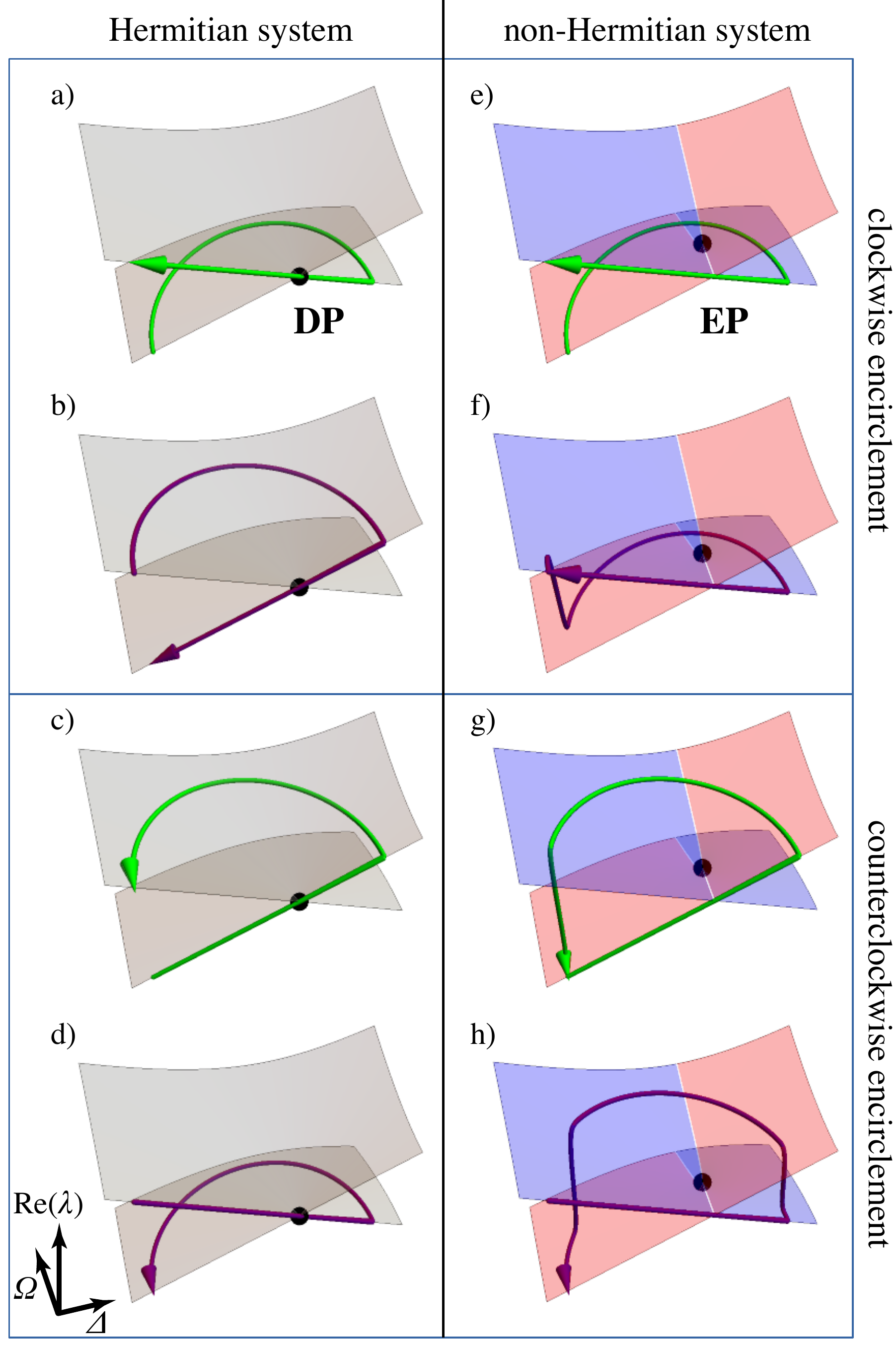}
\caption{Clockwise (top panels) and counterclockwise (bottom panels) passage along a semicircular loop crossing the DP (left panels) and encircling the EP (right panels). The arrows show a projection of the evolving state onto the real part of the eigenspectrum according to Eq.~(\ref{visu}). Violet and green arrows show the state evolution starting at the first and second level, respectively. Red (gain) and blue (loss) regions represent the eigenvalues with $\mbox{Im} \lambda_{\pm{}} > 0$ and $\mbox{Im} \lambda_{\pm{}} < 0$, respectively.
} \label{Fig-1p}
\end{figure}
\section*{Dynamical evolution: Hermitian versus non-Hermitian system}
In this section we analyze the dynamic evolution of the symmetric switch inherent in RAP and the asymmetric state transfer connected to encircling of an EP. We assume the general Hamiltonian 
\begin{equation}
    \label{h2x2SOM}
\mathcal{H} = \frac{1}{2}
    \begin{bmatrix}
          -\Delta - i \gamma & \Omega \\
              \Omega & \Delta + i \gamma
        \end{bmatrix},
\end{equation}
with complex eigenvalues $\lambda_{\pm} = \pm \lambda$, where $\lambda = \sqrt{(\Delta + i \gamma)^2 + \Omega^2}/2\,$, and right eigenstates $\mathcal{H} \vec{r}_{\pm{}} = \lambda_{\pm{}} \vec{r}_{\pm{}}$ defined as
\begin{equation}
    \label{eig}
\vec{r}_{-{}} = 
    \begin{pmatrix}
          \cos \vartheta/2  \\
              \sin \vartheta/2
        \end{pmatrix}, \;\;
\vec{r}_{+{}} = 
    \begin{pmatrix}
          -\sin \vartheta/2  \\
              \cos \vartheta/2
        \end{pmatrix},   
\end{equation}
where $\vartheta$ satisfies $\tan \vartheta = - \Omega / (\Delta + i \gamma)$. Then the solution of the \Schro\ equation $i \partial \vec{\psi}(t) / \partial t = \mathcal{H}(t) \vec{\psi}(t)$ can be expanded in the basis of the instantaneous eigenvectors in the form
\begin{equation}
    \label{eigg}
\vec{\psi}(t) = c_{-}(t) \vec{r}_{-}(t) + c_{+}(t)  \vec{r}_{+}(t),
\end{equation}
where $c_{\pm{}}(t)$ are the complex amplitudes of the state vector in the instantaneous eigenbasis.

To visualise the dynamical evolution along the semicircular loop in the Hermitian and non-Hermitian system we project the evolving state onto the real part of the eigenspectrum. The corresponding trajectories with the vertical coordinate defined as
\begin{equation}
    \label{visu}
\frac{\Re\left[\lambda_{+}(t)\right]\abs{c_{+}(t)}^2 + \Re\left[\lambda_{-}(t)\right]\abs{c_{-}(t)}^2}{\abs{c_{+}(t)}^2 + \abs{c_{-}(t)}^2 }
\end{equation}
are shown in Fig.\ \ref{Fig-1p}. The left column shows the dynamic evolution in the Hermitian system ($\gamma = 0$) along the closed semicircular loop crossing the diabolic point (DP). In CW as well as in CCW direction the eigenstates interchange symmetrically confirming successful RAP. The right column then shows the evolution in the non-Hermitian system along the same semicircular parametric loop which now encircles the EP. As a result of the occurring sudden non-adiabatic jumps between the loss (blue) and gain (red) parts of the eigenspectra, the final states at the end of the loop depend only on the encircling direction and are independent of the initial state confirming the chiral state transfer (\idest asymmetric switching).

\section*{Crossover between symmetric and asymmetric switching}
\begin{figure}[!tb]
\centering
\includegraphics[clip,width=0.90\linewidth]{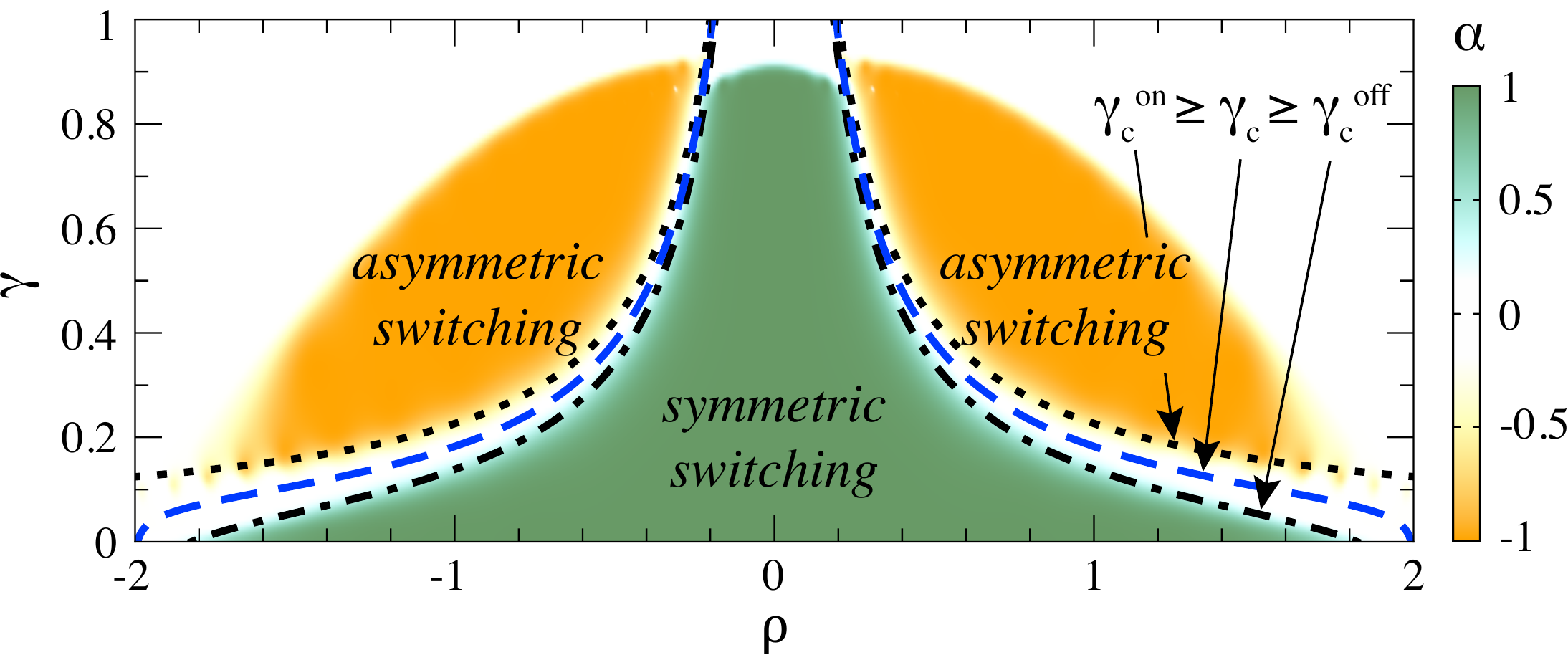}
\caption{The same map as in the main text depicting the state switch asymmetry when numerically following $SC_{A \rightarrow B}$ in both encircling directions as a function of the loop offset $\rho$ and the loss-gain value $\gamma$. The shown switching parameter $\alpha$ takes on its limiting values $1$ ($-1$) for a symmetric (asymmetric) switch as in RAP (as in the chiral state transfer). The two boundaries $\gamma_c^{\toff}$ (black dot-dashed line) and $\gamma_c^{\ton}$ (black dotted line) demarcate the breakdown of the symmetric state transfer and the onset of the asymmetric switching behavior, respectively. In the limit of quasi-adiabatic passage, \idest $T \rightarrow \infty$, those boundaries converge towards $\gamma_c$ shown as a blue dashed line. 
} \label{Fig-Alex}
\end{figure}

In the previous section we have shown that adding a suitable amount of loss to RAP schemes can turn the symmetric state transfer into an asymmetric one. The goal in this section is to determine the critical loss contrast $\gamma_c$ that has to be added to a semicircular loop such that the system then exhibits an asymmetric switching. As it turns out, for finite loop times $T$ there are in fact two boundaries ($\gamma_c^{\toff}$ and $\gamma_c^{\ton}$) that converge towards $\gamma_c$ [Eq.~(6) in the main text] in the limit $T \rightarrow \infty$ [see Fig.~\ref{Fig-Alex}]: the two encircling directions independently switch their behavior, such that at first an additional non-adiabatic jump in one direction turns off the symmetric state transfer at $\gamma_c^{\toff}$. When the loss contrast is increased further to $\gamma_c^{\ton} \geq \gamma_c^{\toff}$, the non-adiabatic jump in the other encircling direction is suddenly inhibited and the overall state transfer becomes asymmetric.  \\
Before we calculate the boundary between the symmetric and asymmetric region, we want to define a measure that allows to quantify the faithfulness of the symmetric and asymmetric state transfer and that highlights the boundaries between those two regimes. For this purpose we firstly combine the values of level population $p$ [Eq.~(5) in the main text] at the beginning and end of the evolution as 
\begin{equation}
    \label{sw}
    S_j = p_j(t=0)p_j(t=T), \ \ \ j = 1,2\; ,
\end{equation}
which equals $-1$ if the eigenstate at the end does not resemble the initial one or $+1$ if the state vector returns back to the initial level. Then considering the state switching in the CW and CCW encircling direction we can define the switching parameter
\begin{equation}
    \label{asw}
    \alpha = \frac{S^{\CW}_1 S^{\CW}_2 + S^{\CCW}_1 S^{\CCW}_2 + S^{\CW}_1 S^{\CCW}_1 + S^{\CW}_2 S^{\CCW}_2}{4}.
\end{equation}
The value $-1$ represents asymmetric switching (chiral state transfer) and $+1$ characterizes symmetric switching. The breakdown of the symmetric region occurs when $0 \leq \alpha \lesssim 1/2$ and the onset of the asymmetric regions occurs if $0 \geq \alpha \gtrsim -1/2$. The particular definition of $\alpha$ allows to distinguish between the breakdown of the symmetric region and the onset of the asymmetric state transfer. When the system is transitioning from a symmetric to an asymmetric state transfer then $\alpha \approx 0$. However, when the loss contrast $\gamma$ becomes too large, the state vector simply ends up in the eigenstate that is subject to gain at the end of the loop for any initial configuration and $\alpha = 0$. The map of the switching parameter $\alpha$ is shown in Fig.~2(e) in the main text as a function of the loop's offset $\rho$ and loss/gain strength $\gamma$.

As can be recognized in Figs.~2(c) and 2(d) in the main text, the collapse of symmetric switching is related to the asymmetry of the CW and CCW evolution of the state initially populating the gain eigenvector (green curves). At the early part of the evolution this eigenstate is amplified and the evolution is adiabatic until the loop crosses the $\Im \lambda = 0$ line at the critical time $t_{*}$ (dashed vertical line), which is different for each encircling direction due to the loop's offset $\rho$. For $t > t_{*}$ the same eigenstate is suddenly attenuated and the adiabatic evolution becomes unstable. The onset of a non-adiabatic jump from the now attenuated towards the instantaneously amplified eigenstate, however, occurs at a delayed time $t = t_{+} > t_{*}$ \cite{MilburnPRBSOM}. For $\rho < 0$ the critical time $t_{*}$ in the CW direction is larger than $T/2$ and hence $t_{+} > T$, which inhibits a non-adiabatic jump for a single passage of the loop. However, for CCW encirclement we have $t_{*} < T/2$ and for a sufficiently large asymmetry $\abs{\rho}$ we get $t_{+} < T$, \idest a non-adiabatic transition occurs, which marks the breakdown of the symmetric state transfer.     

To derive an analytical formula for the border between the regions of symmetric and asymmetric switching we utilize the formalism of stability loss delay described in detail in \cite{MilburnPRBSOM}. In accord with the expansion in Eq.~(\ref{eigg}) we start by introducing the time evolution operator $\mathcal{U}$ defined via $\vec{\psi}(t) = \mathcal{U}(t)\vec{\psi}(0)$ with
\begin{equation}
    \label{u2x2}
    \dot{\mathcal{U}} = -i 
    \begin{bmatrix}
          -\lambda(t) & -f(t) \\
              f(t) & \lambda(t)
        \end{bmatrix} \mathcal{U},\ \ \ \ 
\mathcal{U} = 
    \begin{bmatrix}
          U_{-,-} & U_{-,+} \\
              U_{+,-} & U_{+,+}
        \end{bmatrix},
\end{equation}
where
\begin{equation}
    \label{efu}
f(t) = \frac{\Omega(t) \dot{\Delta}(t) - (\Delta(t) + i \gamma)\dot{\Omega}(t)}{8 i \lambda^2(t)}
\end{equation}
is the non-adiabatic coupling of the eigenstates. Then we define the non-adiabatic transition amplitude
\begin{equation}
    \label{rfu}
R(t) = \frac{U_{-,+}(t)}{U_{+,+}(t)} , 
\end{equation}
which resembles adiabaticity of the dynamic evolution starting from the state populating solely the eigenvector $\vec{r}_{+}$. If $\abs{R} \ll 1$ the state is evolving adiabatically while for $\abs{R} \gg 1$ a non-adiabatic jump has occurred during the evolution. The non-adiabatic transition amplitude $R(t)$ is a solution of the nonlinear differential equation 
\begin{equation}
    \label{rfuc}
\dot{R}(t) = 2 i \lambda(t) R(t) + i f(t) \left[1 + R(t)^2\right],
\end{equation}
with the initial condition $R(0) = 0$. The solution to Eq.~(\ref{rfuc}) follows one of two fixed points with fast non-adiabatic transitions between them. The fixed points are well approximated by 
\begin{equation}
    \label{rfuca}
R^{\ad}(t) \simeq -\frac{f(t)}{2 \lambda(t)}, \ \ \  R^{\nonad}(t) \simeq -\frac{2 \lambda(t)}{f(t)},
\end{equation}
with $\abs{R^{\ad}(t)}  \ll 1$, $\abs{R^{\nonad}(t)} \gg 1$ and $R^{\ad}(t) R^{\nonad}(t) = 1$.  Therefore, the time $t_{+}$ corresponding to the position of a non-adiabatic jump can be determined by the condition
\begin{equation}
    \label{conda}
\abs{R(t_{+})} = 1.
\end{equation}
As long as the loss contrast $\gamma$ is sufficiently small the solution $R(t)$ to Eq.~(\ref{rfuc}) will simply follow $R^{\ad}(t)$. Upon the increase of $\gamma$ a non-adiabatic transition will set in, which demarcates the breakdown of the symmetric switching behavior. To pinpoint the exact location of this boundary we define the critical loss contrast $\gamma_c$ such that the non-adiabatic transition happens exactly at the end of the parameter path, \idest 
\begin{equation}
    \label{condaT}
\abs{R(t_{+} = T)} = 1. 
\end{equation}
At first, we require a suitable approximation for $R(t)$ that correctly reproduces the position of the non-adiabatic jump. We consider initially a stable adiabatic evolution where $R(t)$ closely follows $R^{\ad}(t)$. Then Eq.~(\ref{rfuc}) can be linearized 
\begin{equation}
    \label{liner}
    \dot{R}(t) = 2 i \lambda(t) R(t) + i f(t)
\end{equation}
with the solution
\begin{equation}
    \label{liners}
R(t) = R(0) e^{\Phi(t)} + i \int_0^t f(t') e^{\Phi(t)-\Phi(t')} dt',
\end{equation}
where
\begin{equation}
    \label{faza}
\Phi(t) = 2 i \int_0^t \lambda(t') dt'.
\end{equation}
Expanding the integral in Eq.~(\ref{liners}) through an $N$-times integration by parts and utilizing the properties of asymptotic series we can rewrite the non-adiabatic transition amplitude in the form 
\begin{equation}
    \label{asymr}
R(t) \simeq  \mathcal{R}^{\ad}(t) + D(t) e^{\Phi(t)-\Phi(t_{*})} + A e^{\Phi(t)},
\end{equation}
where
\begin{equation}
    \label{adas}
\mathcal{R}^{\ad}(t) = \sum_{n = 0}^{N - 1} \left( \frac{1}{2 i \lambda(t)} \frac{d}{dt} \right)^n R^{\ad}(t)
\end{equation}
is an optimally truncated correction to $R^{\ad}(t)$, $D(t)$ is the remaining part of the solution not included in the sum in Eq.~(\ref{adas}) and 
\begin{equation}
    \label{ackoop}
A = R(0) - \mathcal{R}^{\ad}(0)
\end{equation}
reflects how the value of $R$ initially differs from the adiabatic fixed point $R^{\ad}$.
The second and third term in Eq.~(\ref{asymr}) are attenuated until $t = t_{*}$ therefore for $t < t_{*}$ the adiabatic term  $\mathcal{R}^{\ad}(t)$ dominates. For $t > t_{*}$, the second and third term in Eq.~(\ref{asymr}) start to grow exponentially and in the vicinity of $t_{+}$ they outgrow the adiabatic term, which leads to the onset of the non-adiabatic transition. To examine the condition Eq.~(\ref{conda}) we are interested in $R$ in the vicinity of $t_{+}$. Therefore, we can neglect the adiabatic term $\mathcal{R}^{\ad}(t)$ in Eq.~(\ref{asymr}). Moreover, as discussed in \cite{MilburnPRBSOM}, in the case of a single passage of the loop the non-adiabatic transition is driven by the third term in Eq.~(\ref{asymr}) since the solution $R$ does not have enough time to approach the adiabatic fixed point $R^{\ad}$ sufficiently closely by the critical time $t_{*}$. Then we can approximate $R$ in the vicinity of $t_{+}$ as
\begin{equation}
    \label{asymrt}
R(t_{+}) \simeq  A e^{\Phi(t_{+})}.
\end{equation}
Since $R(0) = 0$ and the sum in Eq.~(\ref{adas}) is well approximated at $t = 0$ by its $0$-th term we can write
\begin{equation}
    \label{ackoepp}
    A = - \mathcal{R}^{\ad}(0) \approx - R^{\ad}(0) \simeq \frac{f(0)}{2 \lambda(0)}.
\end{equation}
This gives us a viable approximation for $R$ in the vicinity of the non-adiabatic transition
\begin{equation}
    \label{asymrtapprox}
R(t_{+}) \simeq  \frac{f(0)}{2 \lambda(0)} e^{\Phi(t_{+})}.
\end{equation}
To identify the critical loss contrast $\gamma_c$ defining the border between the regions of symmetric and asymmetric evolution, we continue by inserting Eq.~(\ref{asymrtapprox}) into the boundary condition [Eq.~(\ref{condaT})] 
\begin{equation}
    \label{rendo}
\abs{ \frac{f(0)}{2 \lambda(0)}  e^{\Phi(T)} } = \abs{ \frac{f(0)}{2 \lambda(0)} } e^{\Re[\Phi(T)]}  = 1.
\end{equation}
Employing the definition of the semicircular loop [Eqs. (3) and (4) in the main text] we obtain the eigenvalues $\lambda$ in the form
\begin{multline}
    \label{eigam}
\lambda(t) = \frac{r}{2}\sqrt{1 + \Gamma^2 + 2 \Gamma \cos(\pi t/T)} \\
= \frac{r}{2} \bigg\{ 1 + \Gamma \cos \left( \frac{\pi t}{T} \right) + \frac{\Gamma^2}{2} \left[1 - \cos^2 \left( \frac{\pi t}{T} \right) \right] \bigg\} + \bigO(\Gamma^3),
\end{multline}
with $\Gamma = (\rho + i \gamma)/r$, where we expanded $\lambda$ around $\Gamma = 0$. Then
\begin{equation}
    \label{phagam}
\Phi(T) = 2 i \int_0^T \lambda(t') dt' \approx i r T \left( 1 + \frac{\Gamma^2}{4}  \right),
\end{equation}
which yields
\begin{equation}
    \label{rephagam}
\Re[\Phi(T)] \approx   - \frac{\rho \gamma T}{2 r}.
\end{equation}
Using Eqs.~(\ref{efu}) and (\ref{eigam}) we get
\begin{equation}
    \label{flam}
\left| \frac{f(0)}{2 \lambda(0)} \right| = \left| \frac{i \pi}{2 T r} \frac{1}{(1 + \Gamma)^2} \right| =  \frac{\pi}{2 T r} \frac{1}{(\gamma/r)^2 + (1 + \rho/r)^2}.
\end{equation}
Finally, we can rewrite the condition from Eq.~(\ref{rendo}) into the form
\begin{equation}
    \label{condaf}
\frac{\pi}{2 T r} \frac{e^{-\frac{\rho \gamma T}{2 r}}}{(\gamma/r)^2 + (1 + \rho/r)^2} = 1.
\end{equation}
Assuming $\gamma \ll r + \rho$ we can obtain the border for the breakdown of the symmetric state transfer in an analytical form
\begin{equation}
    \label{gamcSOM}
    \gamma_c^{\toff} \simeq \frac{2 r}{T \rho} \ln \left[ \frac{\pi r}{2 T (\rho + r)^2} \right].
\end{equation}
In the derivation of Eq.~(\ref{gamcSOM}) we assumed $\rho < 0$. When $\rho > 0$ the non-adiabatic transition that determines the end of the symmetric switch occurs in the opposite encircling direction. The procedure for the approximation of the critical loss rate is analogous though and so the formula that is valid for all values of $\rho$ is
\begin{equation}
    \label{gamcabs}
    \gamma_c^{\toff} \simeq \frac{2 r}{T |\rho|} \ln \left[ \frac{2 T (r - |\rho|)^2}{\pi r} \right].
\end{equation}
As mentioned before, this particular loss contrast solely determines the point at which the system does not show a symmetric switch in both directions anymore. However, the encircling direction in which the $\Im \lambda = 0$ line is crossed later in time still shows the symmetric state transfer although the final state can have a considerable non-adiabatic contribution. \\
In this regard, Eq.~(\ref{gamcabs}) solely specifies the onset of a non-adiabatic transition in one direction. The derivation in the opposite direction follows the same procedure and the critical loss contrast for the onset of the asymmetric switching behavior can be obtained by simply setting $r \rightarrow -r$ and $T\rightarrow -T$, which reverses the parameter path. The critical loss turns out to be 
\begin{equation}
    \label{gamcabs2}
    \gamma_c^{\ton} \simeq \frac{2 r}{T |\rho|} \ln \left[ \frac{2 T (r + |\rho|)^2}{\pi r} \right].
\end{equation}
It holds that $\gamma_c^{\ton} \geq \gamma_c^{\toff}$ where the equality only hold in the limit $T \rightarrow \infty$. The value at which both of them converge when the loop time $T$ is increased is their mean value 
\begin{equation}
    \label{gamcabsmean}
    \gamma_c \simeq \frac{2 r}{T |\rho|} \ln \left[ \frac{2 T (r^2 - \rho^2)}{\pi r} \right],
\end{equation}
shown as a blue dashed line in Fig.~2(e) in the main text. In Fig.~\ref{Fig-Alex} those three boundaries $\gamma_c^{\ton} \geq \gamma_c \geq\gamma_c^{\toff}$ are drawn on the same map of the switching parameter as in Fig.~2(e) in the main text. 

\section*{Hamiltonian of the waveguide with absorber}
The procedure of how to convey the temporal evolution of a quantum state driven by a $2\!\times\!2$-Hamiltonian to the spatial distribution of microwaves along a bimodal waveguide was described and derived in detail in \cite{DopplerNatureSOM}. First, we briefly summarize the main ideas of this process. Then, in addition to the results presented in \cite{DopplerNatureSOM}, we apply the model to the waveguide with the continuous position-dependent absorber in order to support the numerical results of microwave transport with the calculations based on the semi-analytical model.  

In the two-dimensional waveguide, the propagation of microwaves with frequency $\omega$ can be described by the state $\varphi(x,y,t) = \phi(x,y) e^{- i \omega t}$  satisfying the Helmholtz equation 
\begin{equation}
    \label{helm}
\Delta \phi(x,y) + \epsilon(x,y) k^2 \phi(x,y) = 0,
\end{equation}
where $k = \omega/c$ and $\epsilon(x,y) = 1 + i \eta(x,y)/k$ is a complex dielectric function with $\eta(x,y)$ describing the losses to the environment and to an absorber located in the waveguide interior.

We study wave transmission through a 2D waveguide with constant (transverse) width $W$ and periodically modulated edges described by a profile $\xi(x) = \sigma \sin(k_b x)$. Hard wall boundary conditions are assumed at $y = \xi(x)$ and $y = W + \xi(x)$. 

The microwave wavefunction in this periodic waveguide can be described through a Bloch wave ansatz
\begin{equation}
    \label{blocho}
\phi(x,y) = \Lambda (x,y) e^{i K x},
\end{equation}
where $K$ is a wave number reduced to the first Brillouin zone and $\Lambda (x,y) = \Lambda (x+l,y)$ is a periodic function with the period of the edge modulation $l = 2 \pi/k_b$. In the case of a straight waveguide ($\sigma = 0$) without losses ($\eta = 0$), $\Lambda (x,y)$ has a simple form
\begin{equation}
    \label{blolam}
\Lambda_{mn}^0 (x,y) = e^{i k_b m x} \sin\left( \frac{\pi n y}{W} \right),
\end{equation}
and the corresponding wave number $K^0 \in [-k_b/2,k_b/2 ]$ is given by
\begin{equation}
    \label{blok}
k^2 = \left( k_b m + K^0 \right)^2 + \left( \frac{\pi n}{W} \right)^2.
\end{equation}
Tuning the waveguide width $W$ and/or the frequency $\omega$ of the microwaves such that $2 \pi /W < k < 3 \pi/W$, we reduce the number of propagating modes to two, \idest $n = 1,2$. 

It is known in wave scattering theory for waveguides with modulated boundaries that when the boundary oscillations are given by $k_b = k_r = k_1 - k_2$, where $k_j = \sqrt{k^2 - (\pi j/W)^2}$, both propagating modes experience resonant forward scattering and backscattering of microwaves is negligible. Therefore, we can assume that when $k_b$ is close to $k_r$, \idest $k_b = k_r + \delta$, where $\delta$ denotes a shift from the forward scattering resonance, the wave is moving only in one direction (e.g. from left to right or vice versa) and $K$ has the same sign for all possible solutions [Eq.~(\ref{blocho})]. Then for a given $\omega$ there are two right-propagating solutions of Eq.~(\ref{helm}) in the straight waveguide given by
\begin{equation}
    \label{two}
\phi_{j} (x,y) = e^{i K^0_j x} \Lambda^0_j(x,y), \quad
\Lambda^0_j(x,y) = e^{i k_b m_j x} \sin\left( \frac{\pi j y}{W} \right),
\end{equation}
where $j = 1,2$, $K^0_j > 0$ and $m_j$ are given by Eq.~(\ref{blok}). Setting $k_b = k_r = k_1 - k_2$ (\idest $\delta = 0$) we get $m_2 = m_1 - 1$ and $K^0_1 = K^0_2 = K^0$, which means that the states from Eq.~(\ref{two}) are degenerate with respect to the Bloch wave number $K$. In the following, we will study how this degeneracy is lifted when introducing a finite (but small) periodic modulation of the waveguide edges parametrized by the amplitude $\sigma$ and shift $\delta$. 

Treating the edge modulations as a small perturbation we can write the perturbed Bloch solution of Eq.~(\ref{helm}) in the form
\begin{equation}
    \label{lamapro}
\phi (x,y) \approx \left(a_1 \Lambda^0_1(x,y) + a_2 \Lambda^0_2(x,y) \right) e^{i (K^0 + s) x},
\end{equation}
where $s$ is a small correction to the Bloch wave number. Following \cite{DopplerNatureSOM}, utilizing a perturbative approach by keeping only the first-order terms in $\sigma$, $\delta$, $\eta$ and $s$, the Helmholtz equation (\ref{helm}) can be rewritten into a pair of algebraic equations for the coefficients $a_1$ and $a_2$. Then, using the substitution 
\begin{align}
    \label{subst}
c_1 (x) &= i \sqrt{k_1} e^{-i (\delta/2 - s) x} a_1, \\
c_2 (x) &= -i \sqrt{k_2} e^{-i (\delta/2 - s) x} a_2,
\end{align}
these algebraic equations can be recast into a \Schro-like equation 
\begin{equation}
    \label{SchrSOM}
i \frac{\partial}{\partial x}
\begin{pmatrix}c_1 \\ c_2 \end{pmatrix}
= H \begin{pmatrix}c_1\\c_2\end{pmatrix},
\end{equation}
where the Hamiltonian describing the microwave transport in the bimodal \hyphenation{wave-guide}waveguide with periodically modulated edges can be written as
\begin{equation}
    \label{Hper}
H = 
        \frac{1}{2} \begin{bmatrix}
          \delta & 2 B \sigma \\
              2 B \sigma & -\delta
        \end{bmatrix}
        - i \frac{\eta_0 k}{2}
        \begin{bmatrix}
          \Gamma_{11} & \Gamma_{12} \\
              \Gamma^*_{12} & \Gamma_{22}
        \end{bmatrix},
\end{equation}
where $B = 2 \pi^2 / W^3 \sqrt{k_1 k_2}$ and
\begin{equation}
    \label{gamdef}
 \Gamma_{nm} = \frac{e^{i \pi (m-n)/2}}{\sqrt{k_n k_m}} \frac{2}{W l} 
  \int_0^{l} \int_0^W \tilde{\eta}(x,y) \sin\left( \frac{n \pi}{W}y \right)  \sin\left( \frac{m \pi}{W}y \right) e^{-i(k_n - k_m )x } dx\, dy
\end{equation}
with $\tilde{\eta}(x,y) = \tilde{\eta}(x + l,y)$ specifying the periodic spatial distribution of the absorber in the waveguide. In the case of homogeneous bulk absorption [$\tilde{\eta}(x,y) = 1$] the Hamiltonian takes a simple form
\begin{equation}
    \label{Hhomo}
H_{\texthom} = 
       \frac{1}{2} \begin{bmatrix}
          \delta & 2 B \sigma \\
              2 B \sigma & - \delta
        \end{bmatrix}
        - i \frac{\eta_0 k}{2}
        \begin{bmatrix}
          \frac{1}{k_1} & 0 \\
              0 & \frac{1}{k_2}
        \end{bmatrix}.
\end{equation}
Using the proper substitution outlined in \cite{MilburnPRBSOM}, the Hamiltonian describing homogeneous bulk absorption in the waveguide [Eq.~(\ref{Hhomo})] is directly comparable with the model Hamiltonian for a two-level system with gain and loss [Eq.(1) in the main text], which we used to demonstrate the connection between RAP and the chiral state transfer. However, as was shown in \cite{DopplerNatureSOM}, a homogeneous absorption drastically attenuates the microwaves inside the waveguide, which makes such a system impractical for experimental realization.    

To overcome this issue and to additionally optimize the performance of the asymmetric switching device the loss contrast between the eigenmodes of the Hamiltonian in Eq.~(\ref{Hper}) has to be maximized. It was proposed in \cite{DopplerNatureSOM} that the optimal position of the absorber is located at the nodes of one eigenmode. Then, this eigenmode is almost unaffected by damping while, on the other hand, the second eigenmode is strongly attenuated since the absorber is placed in the vicinity of its maxima. 
However, since those nodes are discrete points concentrating the absorption only in the nodes would cause significant backscattering of microwaves. Therefore, in the semi-analytical model from \cite{DopplerNatureSOM} the absorption smoothly changes in the vicinity of the nodes modelled by Gaussian peaks with a finite width sufficient enough to minimize backscattering. 

Due to mechanical and material limitations of realistic absorbers it is very challenging to realize the above concept experimentally. Moreover, it would be necessary to locate the nodes with very high precision which by itself is a very difficult task. Therefore, in the numerical and experimental setup in \cite{DopplerNatureSOM}, the authors used a continuous thin absorber that was placed such that it interpolates between the nodes. This leads to some parasitic damping of the eigenmode that one wanted to keep free of attenuation, but it did otherwise not affect the results studied there. 

The numerical and experimental results of microwave transport in a waveguide with a continuous position-dependent absorber \cite{DopplerNatureSOM} confirm the successful asymmetric switching. What has been missing so far, however, is a semi-analytical model based on the \Schro\ equation (\ref{SchrSOM}) describing the transmission through such a waveguide. In the next subsections we introduce the Hamiltonian for a waveguide with a continuous position-dependent absorber and compare the spatial evolution of microwaves driven by this Hamiltonian with the numerical simulation of the microwave transport based on the method of recursive Green's functions.

\subsection*{Potential of the continuous position-dependent absorber}

\begin{figure}[t]
\centering
\includegraphics[clip,width=1.0\linewidth]{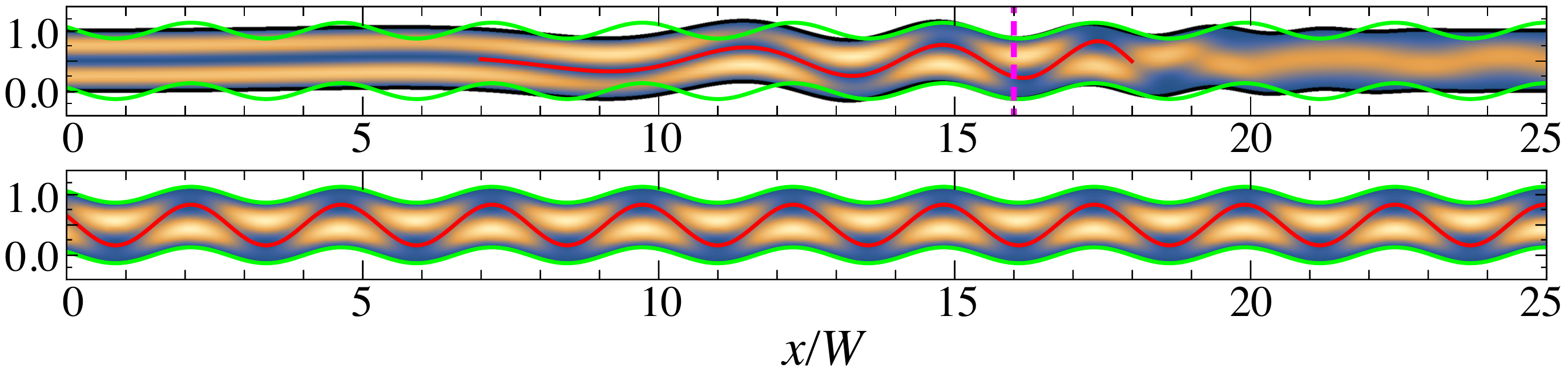}
\caption{Top: Wave density $\abs{\phi(x,y)}^2$ for $k W /\pi = 2.6$ in the finite waveguide without losses calculated semi-analytically using the Hermitian Hamiltonian Eq.~(\ref{Hloss}). The red curve represents the position of the absorber interpolated between the nodes of the wave density. Bottom: Wave density in the infinite periodic waveguide with $\sigma/W = 0.13$ and $\delta W = 0.15$ corresponding to the finite waveguide (top panel) at $x_0 = 16W$ (magenta dashed line). Again, the red curve represents the position of the absorber in the infinite waveguide.} \label{Fig-22}
\end{figure}

As discussed above, the position of the thin continuous absorber is taken from an interpolation between the nodes of one eigenfunction of the Hermitian (lossless) part of the Hamiltonian in Eq.~(\ref{Hper}), \idest
\begin{equation}
    \label{Hloss}
H_0 = 
        \frac{1}{2} \begin{bmatrix}
          \delta & 2 B \sigma \\
              2 B \sigma & -\delta
        \end{bmatrix}.
\end{equation} 
The corresponding Bloch eigenfunctions are given by Eq.~(\ref{lamapro}) with coefficients $a_1$ and $a_2$ related to the eigenvectors of Eq.~(\ref{Hloss}) via Eq.~(\ref{subst}). This yields
\begin{equation}
    \label{coeffs}
a_2 = -i \sqrt{\frac{k_1}{k_2}} \frac{\delta + (-1)^j \sqrt{\delta^2 + 4 B^2 \sigma^2}}{2 B \sigma} a_1, \ \ j = 1,2
\end{equation}
for the $j$-th eigenvector of (\ref{Hloss}).
Then, there are two nodes of the Bloch eigenfunctions located in the unit cell of the periodic waveguide which are periodically distributed along the waveguide at
\begin{gather}
    \label{nodes}
x_o = \pm l/4 + l o, \ \ o \in \mathbb{Z} \\
y_o = \frac{W}{\pi} \arccos{\left( \pm (-1)^j \frac{\abs{a_1}}{2 \abs{a_2}} \right)}.
\end{gather}
As will be clear later we are interested in the node positions of the eigenfunction which is for $\delta < 0$ and negligible amplitude of the edge oscillations $\sigma$ almost entirely equal to $\Lambda^0_2(x,y)$. Therefore, in the next text we assume $j = 2$.
To interpolate between the nodes we use a sine function of the form
\begin{equation}
    \label{posabs}
 y_{int} = \frac{W}{2} + u \sin\left( \frac{2 \pi x}{l} \right),
\end{equation}
centered along the longitudinal axis of the waveguide with amplitude 
\begin{equation}
\label{ampl}
 u = \frac{W}{\pi} \arccos{\left( \frac{\abs{a_1}}{2 \abs{a_2}} \right)-\frac{W}{2}}.
\end{equation}
In the case of a straight waveguide ($\sigma = 0$) one of the eigenfunctions is given by $\phi_2(x,y)$, which means $a_1 = 0$. In this case, the amplitude $u$ is zero and the absorber is simply placed parallel to the center longitudinal axis of the waveguide, which corresponds to the node of the second propagating eigenstate. An example of a waveguide with nonzero $\sigma$ is depicted in the lower panel of Fig.~\ref{Fig-22}. The lower panel shows the wave density $\abs{\phi(x,y)}^2$ for $k W /\pi = 2.6$ in an infinite periodic waveguide with boundary parameters $\sigma/W = 0.13$ and $\delta W = 0.15$. The red curve marks the position of the periodic absorber given by Eq.~(\ref{posabs}) interpolating between the nodes of the depicted eigenfunction of Eq.~(\ref{Hloss}).
Then, the potential of the thin continuous absorber with width $d$ is defined as
\begin{equation}
    \label{potabs}
 \tilde{\eta}(x,y) =  \Htheta\!\left[ y - \frac{W}{2} - u \sin \left( \frac{2 \pi x}{l} \right) + \frac{d}{2} \right] + \Htheta\!\left[  \frac{W}{2} + u \sin \left( \frac{2 \pi x}{l} \right) + \frac{d}{2} -y \right] ,
\end{equation}
where $\Theta$ is the Heaviside step function. Inserting Eq.~(\ref{potabs}) into Eq.~(\ref{gamdef}) results in a Hamiltonian [Eq.~(\ref{Hper})] for the modulated waveguide with the position-dependent continuous absorber. 

\section*{Finite waveguide with position-dependent edge modulation}
We have shown that the unidirectional microwave transport in the periodic bimodal waveguide with modulated edges can be mapped onto the evolution of a quantum state, comprised of the complex amplitudes of the microwaves, that evolve according to the \Schro\ equation (\ref{SchrSOM}) with fixed edge modulation amplitude $\sigma$ and period $\delta$ as well as absorption strength $\eta$. Since we are interested in the dynamical evolution of the state driven by a Hamiltonian with analogous time-dependent parameters, we define the microwave system such that the parameters in the Hamiltonian vary along the longitudinal coordinate $x$. Such a system is realized as a waveguide with finite length $L \gg l$ where the parameters $\sigma(x)$, $\delta(x)$ and $\eta(x)$ vary negligibly slowly on the scale of the edge modulation period $l$. Then, the microwave transport in such a finite waveguide can be well described by the \Schro\ equation (\ref{SchrSOM}) with position-dependent parameters. 
\begin{figure}[!t]
\centering 
\includegraphics[clip,width=0.80\linewidth]{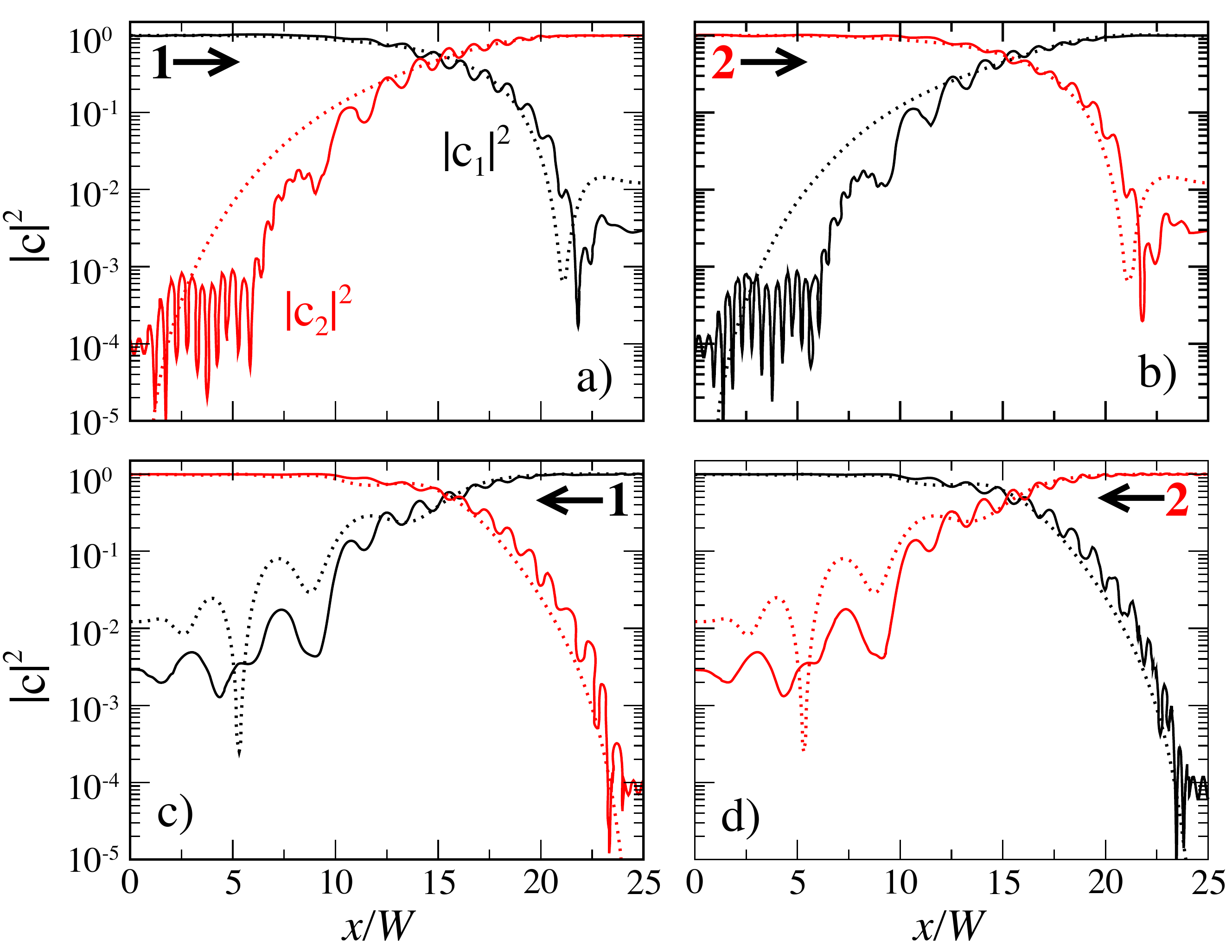}
\caption{Numerical (solid curves) and semi-analytical (dotted curves) calculation of the modal intensities $|c_1|^2$ (black) and $|c_2|^2$ (red) evolving along the waveguide without absorber. Graphs a) and c) depict the wave entering the waveguide in the first mode from the left and right, respectively. Graphs b) and d) show the same for the wave initially in the second mode. All graphs confirm the successful flip of the mode populations characteristic for RAP.} \label{RAP}
\end{figure}
As discussed in the main text, in order to achieve faithful RAP as well as an asymmetric state flip, the variation of the parameters in the Hamiltonian should correspond to a (closed) path which crosses the DP (in the Hermitian case) and encircles an EP (in the non-Hermitian case). The modes at the beginning and end of the evolution are uncoupled which translates to $\sigma = 0$ at both ends of the waveguide. We choose
\begin{equation}
    \label{sigmaline}
\sigma(x) =  \sigma_0 \left[1 - \cos (2 \pi x/L)\right]
\end{equation}
to smoothly increase and decrease the amplitude in order to reduce backscattering of microwaves at both waveguide ends. 
We define the detuning as a linear function of $x$ in the form
\begin{equation}
    \label{deltaline}
\delta(x) =  \pm \delta_0 (2x/L - 1) + \rho,
\end{equation}
where the sign corresponds to traversing the loop in CW or CCW direction, respectively.
The waveguide with parameters $\sigma_0/W = 0.16$, $\delta_0 W = 1.25$, $\rho W = -1.8$ and $L/W = 25$ used in our numerical simulations and in the experiment is depicted in the top panel of Fig.~\ref{Fig-22}.

\begin{figure}[!tb]
\centering
\includegraphics[clip,width=0.80\linewidth]{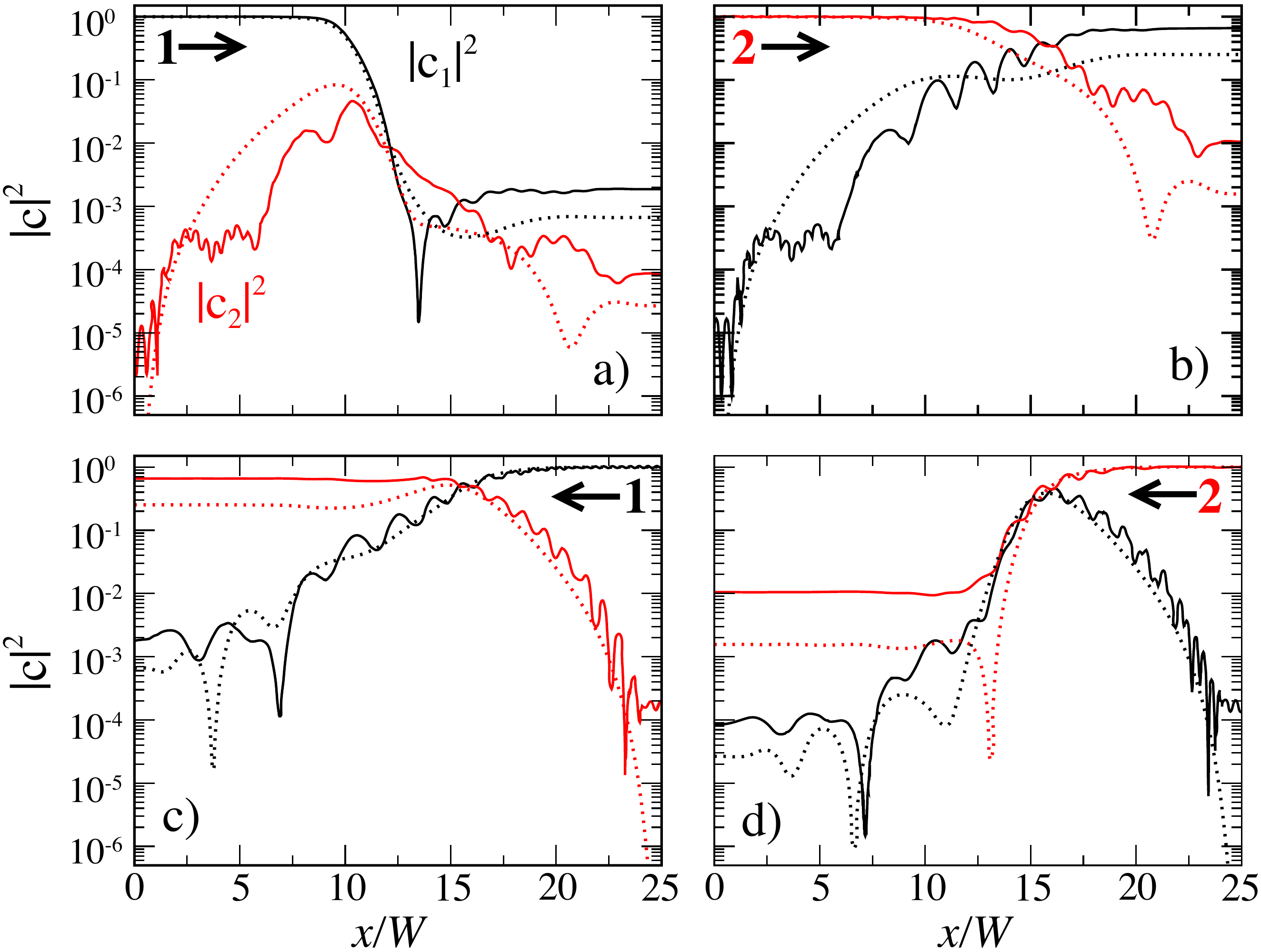}
\caption{Numerical (solid curves) and semi-analytical (dotted curves) calculation of modal intensities $|c_1|^2$ (black curves) and $|c_2|^2$ (red curves) evolving along the waveguide with position-dependent continuous absorber. Graphs a) and c) depict the wave entering the waveguide in the first mode from the left and right respectively. Graphs b) and d) show the same for the wave initially in the second mode. Entering the waveguide from the left the resulting wave intensity is almost entirely composed of the first mode, independently on the initial wave configuration. Entering the waveguide from the right the resulting wave intensity is mostly composed from the second mode. This behaviour confirms the successful asymmetric switching.} \label{CHIR}
\end{figure}

However, in the finite waveguide with varying edge modulations the value of the detuning $\delta(x_0)$ at an arbitrary $x = x_0$ is not exactly the parameter that enters the Hamiltonian in Eq.~(\ref{Hper}), as it was derived for an infinite waveguide with periodic edge modulations. As described in \cite{DopplerNatureSOM}, to obtain the correct value of detuning, the phase $\alpha(x) = [k_r + \delta(x)]x$ of the boundary $\sigma \sin \left[ \alpha(x) \right]$ defining the edge of the finite waveguide has to be linearized, i.e. the edge of the infinite waveguide corresponding to $x = x_0$ is defined as $\sigma \sin[\beta(x)]$ where
\begin{equation}
    \label{betak}
\beta(x) = \left( \frac{d \alpha}{dx}\bigg|_{x = x_0} \right) x = \left[k_r + \Delta(x_0)\right]x 
\end{equation}
and the renormalized position-dependent detuning entering the Hamiltonian of Eq.~(\ref{Hper}) reads
\begin{equation}
    \label{Deltaline}
\Delta(x) = \frac{d (\delta(x) x)}{dx} = \pm \Delta_0 (2x/L - 1) + \rho',
\end{equation}
where $\Delta_0 = 2 \delta_0$ and $\rho' = \delta_0 + \rho$ are the renormalized detuning and offset, respectively.
As an example, the contours of the infinite waveguide corresponding to the finite waveguide defined by Eqs.~(\ref{sigmaline}) and (\ref{deltaline}) at the position $x_0 = 16W$ are shown as green curves in the upper panel of Fig.~\ref{Fig-22}. As expected, in order to achieve successful RAP the renormalized detuning $\Delta$ defined in Eq.(\ref{Deltaline}) is swept through the forward scattering resonance at $\Delta = 0$.

Then, the Hamiltonian describing the microwave transport in the finite waveguide can be written as
\begin{equation}
    \label{wgh2x2}
\mathcal{H} =
    \frac{1}{2}\begin{bmatrix}
          -\Delta & \Omega \\
               \Omega & \Delta
        \end{bmatrix} -
        i \eta \begin{bmatrix}
          \Gamma_{11} & \Gamma_{12} \\
              \Gamma^*_{12} &  \Gamma_{22}
        \end{bmatrix},
\end{equation}
with $\Omega(x) =  2 B \sigma(x)$. In the theoretical calculations and the experimental realization, we locate the absorber in the finite waveguide in the interval $7W < x < 18W$. To reduce the undesired backscattering, the strength (thickness) of the absorber smoothly fades in and out at both ends described by the function
\begin{equation}
    \label{apopin2}
 \eta(x) = \begin{cases} 
      \frac{\eta_0}{4} \bigg\{ 1 - \cos\left[ \frac{2 \pi (x - 7W)}{11 W} \right] \bigg\}^2, & 7 W \leq x \leq 18 W \\
      0\, , & \textrm{ elsewhere} .\\
   \end{cases} 
\end{equation}
In our following semi-analytical calculations, the width of the absorber is $d = 0.019 W$ and the absorption strength $\eta_0 W = 61$.
Driven by the Hamiltonian from Eq.~(\ref{wgh2x2}), the theoretically calculated evolution of the amplitudes $c_1$ (black lines) and $c_2$ (red lines) of the first and second propagation mode ($k W /\pi = 2.6$) along the waveguide is shown in Fig.~\ref{RAP} for the empty waveguide and Fig.~\ref{CHIR} for the waveguide with absorber. Panels a) and c) depict the wave entering the waveguide in the first mode from the left and right, respectively. Graphs b) and d) show the same for the wave initially in the second mode. Arrows mark the direction of the wave propagation. Solid curves denote the results from numerical simulations based on the recursive Green's function method and the dotted curves correspond to the semi-analytical calculation based on the \Schro\ equation (\ref{SchrSOM}) with the Hamiltonian from Eq.~(\ref{wgh2x2}). 

In the Hermitian case [Fig.~\ref{RAP}] the population of the modes almost perfectly flips during the propagation for both encircling directions, which demonstrates faithful RAP. In the non-Hermitian case [Fig.~\ref{CHIR}] the right propagating waves end up almost entirely in the first mode for arbitrary initial wave configurations. On the other hand, the left propagating waves end up almost entirely in the second mode. This clearly proves a successful asymmetric switching.    
Both figures confirm that the results of the semi-analytical model agree very well with the results of the numerical simulation.

The values of $c_1$ and $c_2$ shown here were used to calculate the population inversion $p$ in Fig.~3(c--f) in the main text. The mode populations from our semi-analytical and numerical calculations shown in Fig.~3(c--f) in the main text nicely reproduce the behavior of the simple model driven by the Hamiltonian (\ref{h2x2SOM}) with level populations shown Fig.~2 in the main text. The only significant difference is observed in Fig.~3(f) in the main text, where the state initially in the first mode (violet curve) evolves adiabatically instead of experiencing two non-adiabatic jumps as observed in Fig.~2(f) in the main text. This difference is caused by the fact that in contrast to the simple model the parameter evolution in the waveguide starts at the $\mbox{Im} \lambda = 0$ line [black dashed line in Fig.~\ref{Fig-222}(b), see next subsection] since there is no absorber present initially \cite{MilburnPRBSOM}. 
Chirality of the evolution is preserved, however, since zero or two non-adiabatic jumps both lead to the same final state.

\section*{Position of the EP}
The Hamiltonian in Eq.~(\ref{wgh2x2}) that drives the microwave transport in a finite waveguide in the presence of a continuous position-dependent absorber is in principle defined only along the specific parametric loop given by Eqs.~(\ref{sigmaline}) and (\ref{Deltaline}). Since successful RAP and asymmetric switching is closely related to the position of the DP and EP with respect to the parameter path, we have to extend the definition of the Hamiltonian [Eq.~(\ref{wgh2x2})] to the entire parameter plane $(\Delta,\Omega)$ in order to locate those points. 
\begin{figure}[t]
\centering 
\includegraphics[clip,width=1.0\linewidth]{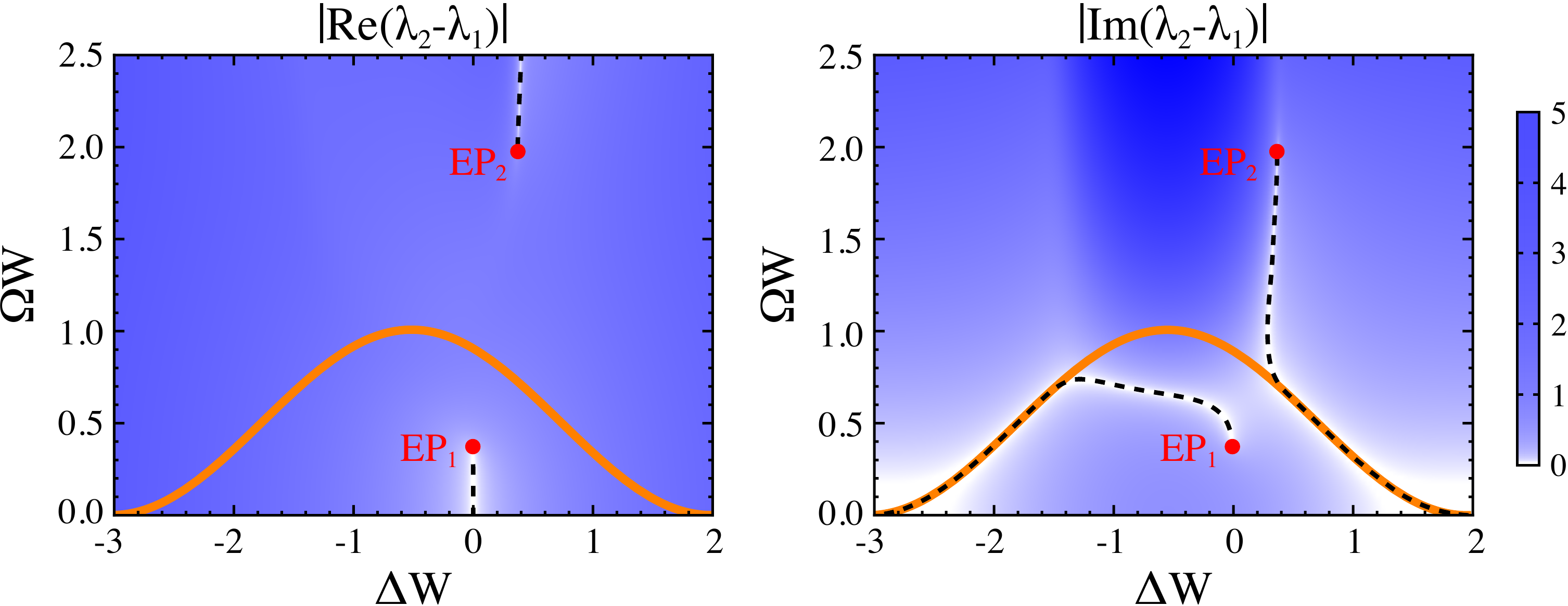}
\vspace{-0.0cm}
\caption{Difference of the real (left panel) and imaginary (right panel) parts of the eigenvalues of the extended non-Hermitian Hamiltonian [Eq.~(\ref{Hamsde})]. Red dots mark the positions of the EPs and the black dashed lines correspond to $\Re \lambda_1 = \Re \lambda_2$ (left panel) and $\Im \lambda_1 = \Im \lambda_2 = 0$ (right panel). The orange curve denotes the parametric loop corresponding to the finite waveguide. The loop starts at the $\Im \lambda = 0$ line and encircles one of the EPs.} \label{Fig-222}
\end{figure}
We choose the extended Hamiltonian in the form 
\begin{equation}
    \label{Hamsde}
\mathcal{H} (\Delta,\Omega) = \mathcal{H}_0(\Delta,\Omega) 
+ i \eta_0 k \Biggr\{ \tilde{\eta}(\Delta) \left[1 - f(\Delta,\Omega)\right] \left[ \frac{\Omega}{\Omega_L(\Delta)} \right]^2
        \begin{bmatrix}
          \Gamma_{11} & \Gamma_{12} \\
              \Gamma^*_{12} & \Gamma_{22}
        \end{bmatrix}
+ \tilde{\eta}_{\texthom}(\Delta) \frac{f(\Delta,\Omega)}{50}
        \begin{bmatrix}
          \frac{1}{k_1} & 0 \\
              0 & \frac{1}{k_2}
        \end{bmatrix}
        \Biggl\},
\end{equation}
where
\begin{equation}
    \label{Ham0}
\mathcal{H}_0 (\Delta,\Omega) = 
        \frac{1}{2}\begin{bmatrix}
         \Delta  & \Omega \\
              \Omega & -\Delta 
        \end{bmatrix} , \ \ 
f(\Delta,\Omega) = \frac{\Omega^2_L(\Delta) - \Omega^2}{\Omega^2_L(\Delta)}
\end{equation}
and
\begin{equation}
    \label{sigmalinek}
\Omega_L(\Delta) = B \sigma_0 \left[ 1 + \cos \left( \pi  \frac{\Delta - \rho'}{\Delta_0} \right) \right].
\end{equation}
The non-Hermitian part of the extended Hamiltonian [Eq.~(\ref{Hamsde})] is an interpolation between the losses due to the continuous thin absorber defined on the parametric loop and the homogeneous absorption for the straight waveguide. The strength of the position-dependent absorber located at $7W < x < 18W$ extended to the parameter plane reads 
\begin{equation}
 \label{etak}
 \tilde{\eta}(\Delta) = 
     \begin{cases}
       \frac{1}{4} \left[ 1 - \cos \left(2\pi \frac{X(\Delta) - 7W}{11W} \right) \right]^2, &\quad 7W < X(\Delta) < 18W \\
       0\, , &\quad\text{elsewhere} \\
     \end{cases}
\end{equation}
with 
\begin{equation}
 \label{etakX}
 X(\Delta) = \left( \frac{\Delta - \rho'}{\Delta_0} + 1 \right)\frac{L}{2}
\end{equation}
and the strength of the homogeneous absorption present in the whole waveguide is
\begin{equation}
 \label{etakt}
 \tilde{\eta}_{\texthom}(\Delta) = \frac{1}{4} \left[ 1 + \cos \left(\pi \frac{\Delta - \rho'}{\Delta_0} \right) \right]^2.
\end{equation}
The DP of the Hermitian part $\mathcal{H}_0$ is simply positioned at $(\Omega_{\DP} = 0,\Delta_{\DP} = 0)$ where the eigenvalues of $\mathcal{H}_0$ coalesce. The locations of the EPs of the non-Hermitian Hamiltonian have to be extracted numerically solving $\lambda_1 = \lambda_2$ where $\lambda_j$ are the complex eigenvalues of $\mathcal{H}$.
The real and imaginary parts of the eigenvalues $\lambda$ of the extended non-Hermitian Hamiltonian [Eq.~(\ref{Hamsde})] are shown in Fig.~\ref{Fig-222}. The orange solid curve corresponds to the parameter loop and the red dots define the position of the EPs. Black dashed curves denote the lines where $\Re\lambda_1 = \Re\lambda_2$ (left panel) and $\Im\lambda_1 = \Im\lambda_2 = 0$ (right panel). As expected, the loop encircles one of the EPs. Moreover, since the absorber is not present at the very beginning and end of the waveguide the eigenvalues are entirely real ($\Im \lambda = 0$) for $x < 7$ and $x > 18$.
Note that there is in principle the freedom to choose the extension of the Hamiltonian arbitrarily which in turn results in different positions for the EP. The only thing that has to be satisfied when the Hamiltonian is extended to the entire ($\Omega, \Delta$)-plane is that the additional Hamiltonian must coincide with the original Hamiltonian along the predefined parameter loop. In fact, the exact position of the EP inside the loop is not important. The crucial point is that the EP is encircled, which can be seen from the topology of the eigenvalues of the Hamiltonian following the parameter loop.

\end{document}